\documentclass[journal]{IEEEtran}
\normalsize

\usepackage{amsmath,amssymb,amsfonts,amsthm}
\usepackage{cite}
\usepackage{graphicx}
\usepackage{textcomp}
\usepackage{algorithm}
\usepackage{algpseudocode}
\usepackage{graphics}
\usepackage{graphicx}
\usepackage{microtype}
\usepackage{subfig}
\usepackage{float}
\usepackage{url}
\usepackage{pdflscape}
\usepackage{supertabular}
\usepackage{enumitem}
\usepackage[table]{xcolor}
\usepackage{multirow}
\usepackage{multicol}
\usepackage{hhline}
\usepackage{acronym}

\def\BibTeX{{\rm B\kern-.05em{\sc i\kern-.025em b}\kern-.08em
    T\kern-.1667em\lower.7ex\hbox{E}\kern-.125emX}}
\begin{document}

\def\sharedaffiliation{%
\end{tabular}
\begin{tabular}{c}}
\title{Interference Suppression Using Deep Learning: Current Approaches and Open Challenges
}
\author{
\IEEEauthorblockN{Taiwo Oyedare$^1$, Vijay K. Shah$^2$, Daniel J. Jakubisin$^{1,3}$ and Jeffrey H. Reed$^1$,~\IEEEmembership{Fellow,~IEEE}}\\
\IEEEauthorblockA{$^1$Dept. of Electrical and Computer Engineering, Virginia Tech, Blacksburg, USA\\
$^2$Dept. of Cybersecurity Engineering, George Mason University, Fairfax, USA \\
$^3$Virginia Tech National Security Institute, Blacksburg, VA, USA (email: djj@vt.edu) \\
\ \{toyedare, djj, reedjh\}@vt.edu, and vshah22@gmu.edu}
}

\maketitle

\begin{abstract}
In light of the finite nature of the wireless spectrum and the increasing demand for spectrum use arising from recent technological breakthroughs in wireless communication, the problem of interference continues to persist. Despite recent advancements in resolving interference issues, interference still presents a difficult challenge to effective usage of the spectrum. This is partly due to the rise in the use of license-free and managed shared bands for Wi-Fi, long term evolution (LTE) unlicensed (LTE-U), LTE licensed assisted access (LAA), 5G NR, and other opportunistic spectrum access solutions. As a result of this, the need for efficient spectrum usage schemes that are robust against interference has never been more important. In the past, most solutions to interference have addressed the problem by using avoidance techniques as well as non-AI mitigation approaches (for example, adaptive filters). The key downside to non-AI techniques is the need for domain expertise in the extraction or exploitation of signal features such as cyclostationarity, bandwidth and modulation of the interfering signals. More recently, researchers have successfully explored AI/ML enabled physical (PHY) layer techniques, especially deep learning which reduces or compensates for the interfering signal instead of simply avoiding it.  The underlying idea of ML based approaches is to learn the interference or the interference characteristics from the data, thereby sidelining the need for domain expertise in suppressing the interference. In this paper, we review a wide range of techniques that have used deep learning to suppress interference. We provide comparison and guidelines for many different types of deep learning techniques in interference suppression. In addition, we highlight challenges and potential future research directions for the successful adoption of deep learning in interference suppression. 
\end{abstract}



\maketitle

\section{Introduction}
\label{sec:introduction}
Interference has always been a dominant and recurring problem in the field of wireless communications. With recent technological breakthroughs such as 5G cellular, internet of things (IoT), LTE-U/LAA, vehicle-to-everything (V2X), and also the increasing availability of low cost software defined radios, coupled with the finite nature of the radio spectrum, interference may continue to be a challenge. With the sheer amount of wireless devices sharing the spectrum band, issues such as homogeneous and heterogeneous technology interference may continue to impact reliable transmission. For instance, not only could legitimate secondary users (SUs) sharing a spectrum band cause interference to one another, rogue SUs could also deliberately interfere with both legacy users as well as legitimate SUs. In Wi-Fi unlicensed applications, Wi-Fi has had to contend with interference from baby monitors, wireless speakers, garage door openers, and more recently, the introduction of unlicensed cellular users in these bands. While earlier interference suppression schemes \cite{kunert2010d1} have utilized interference avoidance schemes usually done in time, frequency, code or space domain, others \cite{mahal2017spectral,khawar2014spectrum,carrick2017optimal} have utilized PHY layer techniques to suppress such interference.  

We consider a broad range of applications where interference could present difficult challenges. These include, for example, automotive radar, non-orthogonal multiple access (NoMA) systems, dense internet-of-things (IoT), full duplex systems, dynamic spectrum use as well as coexistence studies between different technologies. Within the context of a communication channel, interference can be broadly categorized as adjacent channel and co-channel interference. The main difference between these two is that the former occurs between adjacent frequency bands while the latter occurs in the same frequency band. In the context of the technology, interference could be categorized as homogeneous technology interference and heterogeneous technology interference. For the former, signals from the same technology interfere with one another while the latter implies interference between different technologies. In addition, full duplex systems that are aimed at improving spectral efficiency also experience self-interference (SI). Often times, SI usually occur when simultaneous transmission and reception occur in a frequency band. Details of these categories can be found in section \ref{interf_xtics}. 

Traditional interference suppression techniques relied on the interference categories to know which approach to follow. These  techniques include PHY layer schemes, blind source separation schemes (BSS) amongst others. PHY layer schemes rely on suppressing interference via optimizing PHY layer design or providing redundancy to account for interference at the receiver \cite{mahal2017spectral,khawar2014spectrum,carrick2017optimal,carrick2019mitigating,batra2006mitigation}. They investigate optimal PHY layer schemes with baseline methods such as heuristic models, optimization techniques, greedy algorithms, amongst others, to suppress interference. For instance, Carrick \textit{et al.} \cite{carrick2019mitigating} investigated interference suppression methods that exploit spectral redundancies in radar and orthogonal frequency division multiplexing (OFDM) signals using time-varying frequency shift filter referred to as \textit{TV-FRESH} filters. The technique in \cite{carrick2019mitigating} utilized a \textit{TV-FRESH} filter to suppress linear frequency modulated radar interference. The main approach for the TV-FRESH technique is to use spectral redundancy to enhance the desired signal or to estimate and subtract the interfering signal and thus leaving only the signal-of-interest (SOI). In addition, an adaptive filter can be used to remove interference from a signal without affecting crucial information \cite{kesteven2005adaptive,mohammed2021interference}. Similar to the TV-FRESH approach, adaptive filters are able to filter or subtract interference from a signal to yield a signal with less interference. One key attribute of adaptive filters is their ability to adjust the filtering criteria based on the type of interference to be suppressed \cite{kesteven2005adaptive,fite1996blind}. Earlier works that utilized these techniques rely on the use of signal properties such as cyclostationary features \cite{kim2007cyclostationary} inherent in the signal, signal envelope \cite{agee1989property,fite1996blind,agee1993simulation} and channel bandwidth \cite{palicot2003new} to identify and suppress interfering signals. As noted by \cite{reed1995optimal}, the major issue with PHY layer techniques that optimize the weights of the filter using estimated time-frequency statistics is that time-varying interference result in time-varying weights which tend to reduce  the effectiveness of conventional adaptive filters such as Least Mean Squares (LMS). Some of the single channel approaches discussed could also be extended to multiple channels. Furthermore, the reliance on domain expertise or specific signal knowledge may result in lack of scalability when considering the variety and the number of wireless devices present in modern wireless environments.




In investigating BSS techniques, Fabrizio et. al. \cite{fabrizio2014blind} proposed a method that estimates interfering waveforms occupying the same channel as the SOI using a generalized estimation of multipath signals (GEMS) algorithm. In BSS, a set of signals are separated from a set of mixed signals with or without information about the mixing process or even the source signals \cite{cao1996general}. Simply put, BSS is used to recover the original component from a mixture signal. The SOI, as the name implies, is the noise-free component of a signal that we seek to analyze or recover at the receiver \cite{pizurica2006estimating}. Although, a lot of works in the literature considered separating audio signals with fewer dimensions, multi-dimensional data have also seen the application of BSS techniques. For instance, principal component analysis (PCA) is an example of a BSS technique that has been used in the image processing domain for dimensionality reduction. For their BSS technique, authors in \cite{fabrizio2014blind} reconstructed interference components at the receiver from the estimated waveform and then subsequently subtracted these components from the data in time domain. Specifically, they leveraged a least-squares suppression technique to uncover synthetic targets from the GEMS algorithm using waveform filtering. The waveform filtering relies on interference signal reconstruction and time-domain coherent subtraction. Details of the GEMS algorithm can be found in \cite{fabrizio2011adaptive}. Some other suppression techniques different from BSS were also studied. For instance, Uysal \cite{uysal2018synchronous} proposed a time domain approach to decompose signals so as to separate the automotive radar interference signal from the SOI. In addition, Li et. al. \cite{li2016dme} first estimated signal parameters (from known interference types) using a time-modulated windowed all-phase discrete Fourier transform (DFT) and then reconstructed the estimated signal and consequently subtracted it from the received signal to obtain the SOI. In general, the major issue with these approaches is that they tend to require very precise detection of where the interference is located as well as its duration and other characteristics. Specifically, we observe that the traditional interference suppression techniques suffer three main limitations: (1) reliance on domain knowledge and the interference characteristics; (2) their inability to generalize and (3) complexity resulting from detecting the precise location of the interference. 

Deep learning based interference suppression techniques, which are model-free,  have been found to sufficiently address the limitations of traditional interference suppression techniques. For instance, while real-time interference modeling may be difficult, a model-free approach, may remove the need for domain knowledge and thus make them more practical in real world deployment. Since the radio frequency (RF) environment is very stochastic in nature, this goal may seem very difficult to attain especially in the wild where the inability to suppress such could negatively impact performance. However, deep learning approaches can learn directly from the data and may not require any prior knowledge to suppress interference. While it it challenging to generate enough datasets for different realizations of the channel environment, we note that because deep learning usually require access to a large repository of data, software defined radios (SDRs) have simplified the data collection process since they are relatively inexpensive and can be used to generate data in different channel scenarios. While this process may take time, it is important to understand that it would eventually benefit the wireless communication research community as deep learning for interference suppression continues to be studied and adopted.

Another key benefit of using deep learning is the ability to learn powerful, non-linear relationships for interference suppression which would be difficult to develop through traditional suppression techniques. Many traditional interference suppression schemes are not robust when the decision boundary is complex and non-linear.  Deep learning techniques are being embraced in the interference suppression community because they can learn complex non-linear decision boundaries. Finally, the issue of scalability for which many traditional suppression techniques are not able to achieve can be done with deep learning.

In this paper, we seek to address the knowledge gap in the literature with respect to the state-of-the-art in DL-based interference suppression. To account for the limitations of DL-based interference suppression techniques, we discuss several challenges, including lack of interpretability, the stochastic nature of the wireless channel, issues with open set recognition (OSR) and challenges with implementation in section \ref{sec:challenges}. To the best of our knowledge, this is the first paper that explores deep learning
based interference suppression techniques for multiple application areas and various interference characteristics. Our investigation show that not only do deep learning techniques reduce computational complexity in comparison to non-deep learning suppression techniques, they also outperform non-DL techniques in various channel environment as well as different wireless communications applications. We believe that this is a very important study since deep learning is gaining attention in wireless communication research for interference suppression. This paper would best serve researchers working on interference suppression issues or the application of deep learning to wireless communications, as well as, wireless spectrum users and decision makers. Our contributions are as follows: 

\begin{itemize}
\item We identify key characteristics and provide a taxonomy of interference-related issues in wireless communication. This includes key features of adjacent channel interference, co-channel interference, homogeneous and heterogeneous technology interference prevalent in wireless communication applications.
\item We review a wide range of techniques that have used deep learning to suppress interference. Techniques include both supervised deep learning (such as CNN, deep neural networks (DNN) and recurrent neural networks (RNN)), unsupervised techniques such as deep autoencoders. We also discuss the different wireless communication applications for which the deep learning techniques have been utilized. 
\item We summarize the benefits and drawbacks of using different DL algorithms for interference suppression. This summary also include the technical approach utilized as well as the benefits and drawbacks associated with them. Table \ref{table:Adv_Diasdv} has been provided to give a high level summary.
\item We highlight challenges associated with using deep learning for interference suppression. While some of these challenges are not unique to DL-based interference suppression, we note that they have the capability of slowing down the adoption of DL in interference suppression study. 
\end{itemize}



\begin{figure*}[!bt]
\centering
\includegraphics[scale=0.585]{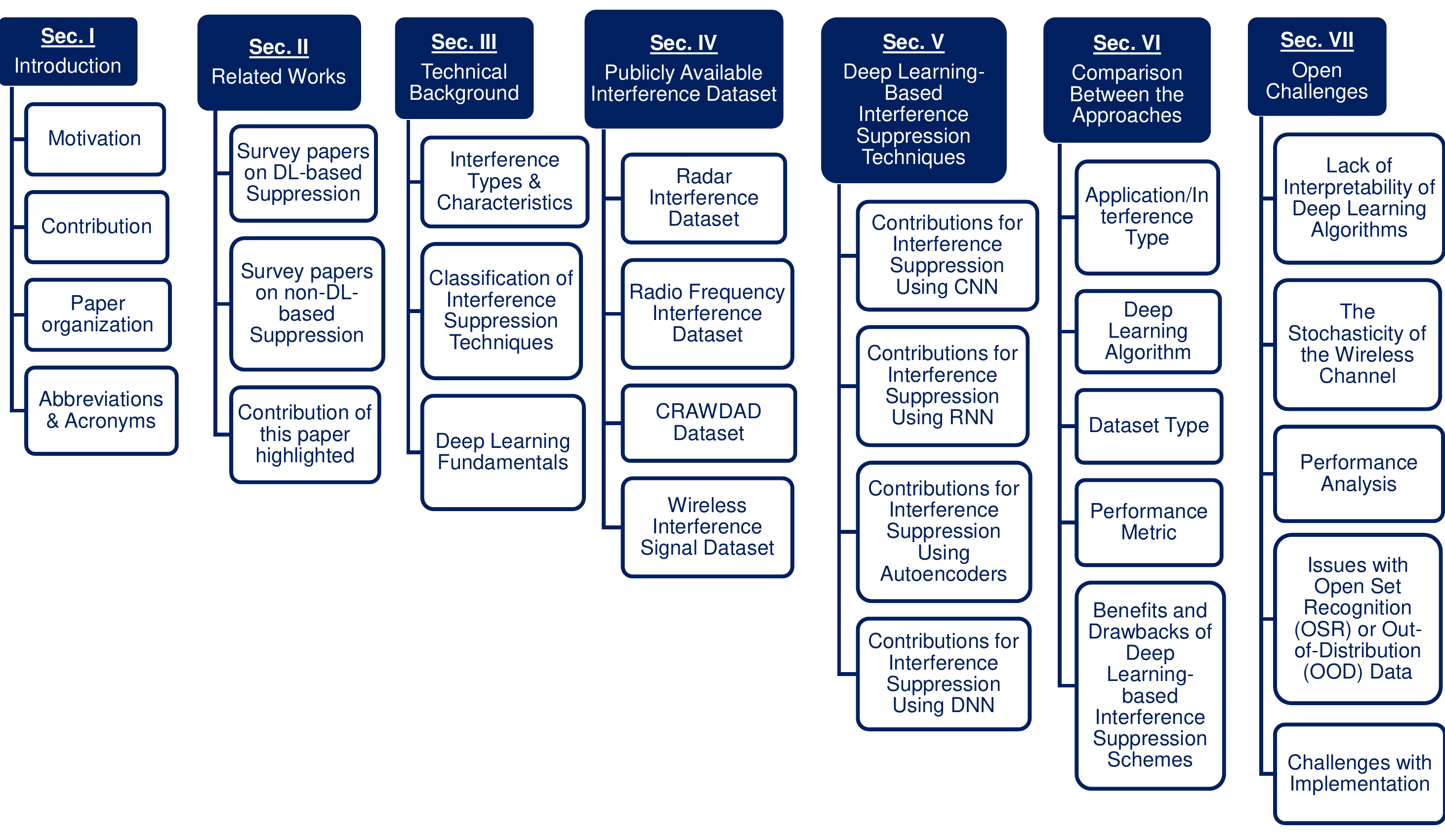}
\caption{High Level Structure of the Survey Paper}
\label{fig:paper_structure}
\vspace{-0.2in}
\end{figure*}

Fig. \ref{fig:paper_structure} shows a high level structure of the paper including a breakdown of the topics discussed in each section. 

\subsection{Abbreviations and Acronyms}
\begin{acronym}[PDP-WPT]
\footnotesize
\acro{ACI}{Adjacent Channel Interference}
\acro{AI}{Artificial Intelligence}
\acro{AUC}{Area Under the receiver operating characteristics Curve}
\acro{AUC}{Area Under the ROC Curve}
\acro{AWGN}{Additive White Gaussian Noise}
\acro{BER}{Bit Error Rate}
\acro{BLE}{Bluetooth Low Energy}
\acro{BPSK}{Binary Phase Shift Keying}
\acro{BSS}{Blind Source Separation}
\acro{CCI}{Co-channel Interference}
\acro{CNN}{Convolutional Neural Network}
\acro{Conv}{Convolution}
\acro{CS}{Chirp Sequence}
\acro{CSMA/CA}{Carrier Sense Multiple Access/ Collision Avoidance}
\acro{CTI}{Cross Technology Interference}
\acro{CVCNN}{Complex Valued Convolutional Neural Network}
\acro{CWI}{Continuous Wave Interference}
\acro{DBP}{Digital Backpropagation}
\acro{DFT}{Discrete Fourier Transform}
\acro{DL}{Deep Learning}
\acro{DNN}{Deep Neural Network}
\acro{DSSS}{Direct Sequence Spread Spectrum}
\acro{EBPNN}{Extended Backpropagation Neural Network }
\acro{EVM}{Error Vector Magnitude}
\acro{FCN}{Fully Convolutional Network}
\acro{FFT}{Fast Fourier Transform}
\acro{FMCW}{Frequency Modulated Continuous Wave}
\acro{GAN}{Generative Adversarial Networks}
\acro{GEM}{Generalized Estimation of Multi-path Signals}
\acro{GPS}{global Positioning System}
\acro{GPU}{Graphical Processing Unit}
\acro{GRU}{Gated Recurrent Unit}
\acro{IMAT}{Iterative  Method  with  Adaptive  Thresholding}
\acro{IoT}{Internet of Things}
\acro{IQ}{In-phase Quadrature}
\acro{LDBP}{Learned Digital Backpropagation}
\acro{LQI}{Link Quality Indicator}
\acro{LR}{Linear Regression}
\acro{LTE-U}{Long Term Evolution Unlicensed}
\acro{MAE}{Mean Absolute Error}
\acro{MCWI}{Multi-tone Continuous Wave Interference}
\acro{MIMO}{Multiple Input Multiple Output}
\acro{MLP}{Multi-layer Perceptron}
\acro{MMSE}{Minimum Mean Square Error}
\acro{MSE}{Mean Square Error}
\acro{MSPE}{Mean Square Prediction Error}
\acro{NoMA}{Non-orthogonal Multiple Access}
\acro{NS3}{Network Simulator 3}
\acro{OFDM}{Orthogonal Frequency Division Multiplexing}
\acro{OOD}{Out of Distribution}
\acro{OSR}{Open Set Recognition}
\acro{OTA}{Over The Air}
\acro{PCA}{Principal Component Analysis}
\acro{PDP-WPT}{Power Distributing Predominance Wavelet Packet Transform}
\acro{PHY}{Physical}
\acro{PPMSE}{Peak Phase Mean Square Error}
\acro{PRR}{Packet Receive Ratio}
\acro{ReLU}{Rectified Linear Unit}
\acro{RFI}{Radio Frequency Interference}
\acro{RFI}{Radio Frequency Interference}
\acro{RNN}{Recurrent Neural Network}
\acro{RSSI}{Received Signal Strength Indicator}
\acro{SCWI}{Single-tone Continuous Wave Interference}
\acro{SDR}{Software Defined Radio}
\acro{SEFDM}{Spectrally Efficient Frequency Division Multiplexing}
\acro{SER}{Symbol Error Rate}
\acro{SINR}{Signal to Noise plu Interference Ratio}
\acro{SIR}{Signal to Interference Ratio}
\acro{SNR}{Signal to Noise Ratio}
\acro{SoI}{Signal of Interest}
\acro{SSFM}{Split-Step Fourier Method}
\acro{SSIM}{Structural Similarity}
\acro{STI}{Same Technology Interference}
\acro{TDMA}{Time Division Multiple Access}
\acro{UKF-RNN}{Uscented Kalman Filter-based Recurrent Neural Network}
\acro{V2X}{Vehicle to Everything}
\acro{VAE}{Variational Autoencoders}
\acro{WBAN}{Wireless Body Area Network}
\end{acronym}

\section{Related Works}\label{sec:related}
To the best of our knowledge, the current literature lacks a comprehensive study on deep learning-based interference suppression. As at the time of writing this paper, we have found only two surveys which consider DL techniques in the context of interference suppression. In \cite{geng2021deep}, the authors discussed deep learning techniques for radar applications in general and briefly highlighted interference techniques that fell into the deep learning categories. One key distinction of our work from theirs is that the interference types they considered includes only jamming and clutter. We note that this investigation is very important and our work goes a step further by investigating more DL-based interference suppression applications including radars, optical communications, IoT networks, GPS navigation, full duplex systems and wireless sensor networks. Further, we consider many other types of interference such as self-interference, adjacent channel interference, homogeneous and heterogeneous technology interference. Furthermore, the jamming suppression techniques considered in \cite{geng2021deep} focused on target signal classification and recognition. Our work is fundamentally different from theirs because we focused more on interference suppression techniques and not interference classification techniques. 
In another work, Chen \textit{et al.} \cite{chen2021coexistence} surveyed both coexistence and interference mitigation techniques, including traditional (non-DL) and DL approaches for wireless personal area network (WPAN) and WLAN applications. While this is also another important contribution, the scope of the application considered is different in comparison to the one presented in this survey. Also, we found that their work focused more heavily on non-DL interference suppression and briefly discussed DL-based suppression techniques. Furthermore, in our own work, we discuss several critical interference characteristics and why DL techniques are suitable to suppress them. Also, we point the community to publicly available dataset for interference suppression. 

Many of the other surveys \cite{azmi2014interference,chowdhury2009interferer,kunert2010d1,lu2006survey,le2015interference} on interference suppression reviewed papers that utilized signal processing techniques and even non-ML techniques to avoid interference. For instance, Azmi et. al. \cite{azmi2014interference} discussed technical papers that used techniques such as transmit power control, dynamic frequency selection, interferer classification schemes \cite{chowdhury2009interferer}, time slotted channel hopping schemes amongst others to suppress interference in wireless sensor networks (WSN). Further, Lu \textit{et al.} reviewed interference mitigation schemes that utilized adaptive array processing in GPS systems. Le \textit{et al.} also surveyed interference mitigation schemes used in wireless body area sensor networks. Kunert et. al. \cite{kunert2010d1} surveyed six basic techniques that researchers have used to suppress radar interference. They sorted the suppression techniques into domains such as polarization, time, frequency, coding and space. While the techniques discussed in this survey paper typically require domain expertise to be successful, it is not clear if they will be scalable in applications such as IoT and 5G where there are multiple devices sharing the same frequency channel. Furthermore, one of the approaches discussed in the paper is "communicate and avoid" in various domains including time and frequency. While this is a straightforward approach, it involves multiple steps to achieve and the time to implement in a latency-sensitive applications such as 5G New Radio (NR) or vehicle to everything (V2X) may be unrealistic. Some of these limitations can be tackled using deep learning techniques surveyed in this paper. 

For automotive radar suppression algorithms, Toth et. al. \cite{toth2019performance} investigated state-of-the-art interference suppression techniques including zeroing, ramp filtering, IMAT and time domain parametric interpolation (IVMip). Their study did not include ML or DL approaches that could be leveraged to suppress interference. Also, their application focused primarily on automotive radar interference, whereas our survey encompasses many other application areas. In the same vein, Nabil et. al. \cite{nabil2015recent}, discussed advances in self and mutual interference suppression techniques for wireless networks. They surveyed redundancy techniques such as MIMO technology to suppress mutual interference in scenarios when the receiver does not have complete knowledge of the interfering signal. They also discussed interference cancellation techniques such as Balun transformer, isolation of the transmitter and receiver paths, phase shifting and spatial techniques (enabling an appropriate distance between the transmit antennas).  

Our work is different from all the other surveys on interference suppression because we explored suppression approaches that utilized deep learning for many application areas and interference characteristic types while also providing a comparison between them and the major challenges with using deep learning to suppress interference. To the best of our knowledge, this is the first paper that explores deep learning based interference suppression techniques for multiple application areas and various interference characteristics. There is a lot of promising results that show that DL based techniques can be very useful in quickly suppressing interference in many wireless communications applications. 

\begin{table*}[!ht]
\caption{Wireless Technologies sharing the $2.4$GHz band with their interference avoidance methods}
\centering
\begin{tabular}{|p{4.5cm}|p{4cm}|p{5cm}|}
\hline
\textbf{Technology} & \textbf{Number of Channels}   & \textbf{Interference Avoidance Method}\\
\hline
Wi-Fi (IEEE 802.11b/g/n/ax) & $3$ $20$MHz static channels & Collision avoidance \\
\hline
Bluetooth & $79$ $1$ MHz dynamic channels & Frequency hopping \\
\hline
Wireless USB (including wireless keyboard, mouse, etc) & $79$ $1$ MHz dynamic channels & Frequency agility \\
\hline
IEEE 802.15.4 & $16$ $5$MHz static channels & Fixed collision avoidance \\
\hline
Bluetooth Low Energy (BLE) & $40$ $2$MHz channels & Frequency hopping \\
\hline
Cordless Phone & $100$KHz (Analog Phone); $800$KHz (Frequency hopping phone) & Frequency hopping \\
\hline
\end{tabular}
\label{table:cti}
\end{table*}

\section{Technical Background}\label{sec:background}

\subsection{Interference Types and Characteristics} \label{interf_xtics}
In this section, we discuss the types and characteristics of interference in the wireless communications research.
\subsubsection{Adjacent Channel Interference}\label{adjacent}
Adjacent channel interference (ACI) occurs when the power from a signal in a channel adjacent to it causes interference to another signal in a different channel. There are several reasons why ACI may occur, for example in frequency modulated (FM) systems, incomplete filtering may lead to ACI. Furthermore, an inadequate frequency control in either the interfering channel or the reference channel may also lead to ACI. In some cases, interference avoidance schemes can be used to avoid ACI. For instance, while the $2.4$~GHz Wi-Fi band supports a total of $11$ channels, it is recommended that Wi-Fi users in the United States utilize non-overlapping channels (take for example: channels $1$, $6$ and $11$) \cite{kajita2016wi}. 

\subsubsection{Co-channel Interference} \label{co-channel}
Co-channel interference (CCI), on the other hand, occurs when two devices or transmitters that share the same frequency for transmission interfere with one another \cite{yang2003co}. In cellular communication applications, CCI is more pronounced when frequency bands are re-used. Interference avoidance schemes alone are  not sufficient to suppress CCI. In this survey, we focus on the more difficult problem of CCI. 

\subsubsection{Self Interference}
Self interference (SI) is a very common phenomenon in full duplex systems where transmitters and receivers utilize the same frequency band for simultaneous communication. In cognitive radio (CR) radio applications, in-band full duplex transmission and reception on the same frequency at the same time is very useful in doubling spectral efficiency while maximizing the use of available spectrum resources \cite{jain2011practical,bharadia2013full}. Techniques such as antenna, analog and digital cancellation have been studied in canceling interference, however, it has been shown that it is very difficult to completely cancel SI due to imperfections such as IQ imbalance, power amplifier non-linearities, amongst others \cite{balatsoukas2018non}.

\subsubsection{Homogeneous Technology Interference}\label{homog}
Homogeneous technology interference, also referred to as same technology interference, occurs when a wireless technology is interfered by a varying number of signals from the same technology. They can occur either in the same channel (co-channel) or in different channels (adjacent channel). In some cases, the number of interfering signals could be more than one, depending on the type of frequency band the dominant signal is utilizing. Such interference is often due to the use of multiple access technologies such as IEEE 802.15.1 and IEEE 802.15.4 \cite{grunau2018multi}. Farid \textit{et al.} \cite{farid2016homogeneous} also studied homogeneous technology interference avoidance technique such as CSMA/CA-based approach; interference suppression techniques such as adaptive techniques, graph coloring, TDMA based approaches, and heuristic techniques, amongst others, for wireless body area networks (WBAN)\footnote{WBAN is a broad term describing wearable sensors connected in a wireless manner to a hub \cite{farid2016homogeneous}.}. Also, code division multiple-access (CDMA), a multiplexing technology for cellular communication technologies, is also vulnerable to homogeneous technology interference. In spread spectrum CDMA, the spread spectrum systems share the band with other communication systems. To do this successfully, individual users in a spread spectrum CDMA system has a unique binary psuedo-noise (PN) code that helps the receiver to distinguish it from the other users in the system. In the past three decades, several interference suppression techniques, such as linear \cite{milstein1988interference} and non-linear estimation techniques \cite{vijayan1990nonlinear}, have been proposed to suppress homogeneous technology interference in CDMA systems. We note that the research in homogeneous technology interference has considerably waned in comparison to heterogeneous technology interference discussed in section \ref{heter}. As a result of this, many of the DL-based interference suppression techniques explored heterogeneous technology interference more than they did homogeneous technology interference. 

\subsubsection{Heterogeneous Technology Interference}\label{heter}
Understanding heterogeneous technology interference, also referred to as cross technology interference (CTI), is very crucial to the coexistence of different wireless technologies sharing similar unlicensed frequency bands. This interference occurs when a wireless technology is interfered by signals from the different technology \cite{grunau2018multi}. Classic examples include licensed LTE vs Wi-Fi \cite{hu2012interference,bayhan2017coexistence}, unlicensed LTE versions\footnote{Unlicensed LTE versions include LTE-Licensed Assisted Access (LTE-LAA), LTE in Unlicensed Spectrum (LTE-U), LTE-WLAN Aggregation (LWA) and Multefire. In general, these unlicensed LTE technologies are either license-anchored or non license-anchored. For license-anchored systems (LTE-U, LAA) the primary carrier (anchor) uses the licensed parts of the spectrum and for non license-anchored systems (Multefire) both control and data traffic are sent through the unlicensed spectrum \cite{naik2018coexistence}.} vs Wi-Fi \cite{naik2018coexistence,chen2016coexistence}, radar\footnote{While radars are the incumbent users of the 5.150-5.350 GHz and 5.470-5.725 GHZ bands, Wi-Fi devices operating in the 5GHz band also share the spectrum when no radar system is transmitting \cite{naik2018coexistence}. This can lead to interference from the Wi-Fi devices to the legacy radar systems.} vs Wi-Fi \cite{tercero2011impact,saltikoff2016threat}, dedicated short range communication (DSRC)\footnote{DSRC supports applications based on vehicular communication \cite{kenney2011dedicated}. The IEEE 802.11p standard, an amendment to the IEEE 802.11 standard, was developed to incorporate the wireless access in vehicular environments (WAVE) vehicular communication system in order to support intelligent transportation systems. This technology operates between 5.85-5.925 GHz band. Since Wi-Fi also operates as a secondary user in the same radar band, they can often times cause interference to the legacy radar technologies.} vs Wi-Fi \cite{naik2017coexistence,lansford2013coexistence}. 

Due to coexistence issues that can lead to interference between these heterogeneous technologies, several studies have proposed several interference suppression techniques \cite{hithnawi2014understanding,hithnawi2016crosszig,wang2021prcomm,liu2020distributed,pulkkinen2020understanding} to better explore the manner in which these technologies interact. In most cases, CTI exhibit distinct recognizable patterns that interference-aware protocols can leverage \cite{hithnawi2014understanding}. Grunau et. al. \cite{grunau2018multi} showed that the impact of  CTI for an IEEE 802.11 b/g signal interfered by IEEE 802.15.1 and IEEE 802.15.1 signals was relatively benign when compared to the converse. This is because both IEEE 802.15.4 and IEEE 802.15.1 signals result in lesser overlapping interference compared to IEEE 802.11 b/g signal because they have narrower bandwidth. However, for low-power wireless devices, CTI can disproportionately impact their coexistence with other devices when sharing the same band~\cite{hithnawi2016crosszig}.


Table \ref{table:cti} shows wireless technologies such as IEEE 802.11 (the standard that defines Wi-Fi), Bluetooth, Bluetooth low energy (BLE), wireless USB, IEEE 802.15.4\footnote{IEEE 802.15.4 is the basis for Zigbee, ISA100.11a, WirelessHART, IPv6 over Low-Power Wireless Personal Area Networks (6LoWPAN) and cordless phone (including analog and frequency hopping phones). Other technologies include microwave oven, wireless camera, baby monitors, car alarms, video devices amongst others. It defines the low-rate wireless personal area networks (LR-WPANs)\cite{gascon2009security}.},  and others sharing the 2.4 GHz band. Despite the fact that these technologies have different power levels and width of the occupied spectrum, that they could still cause interference to one another given the fact that they share the congested unlicensed 2.4 GHz band. 



\begin{figure*}[!ht]
\centering
\includegraphics[scale=0.70]{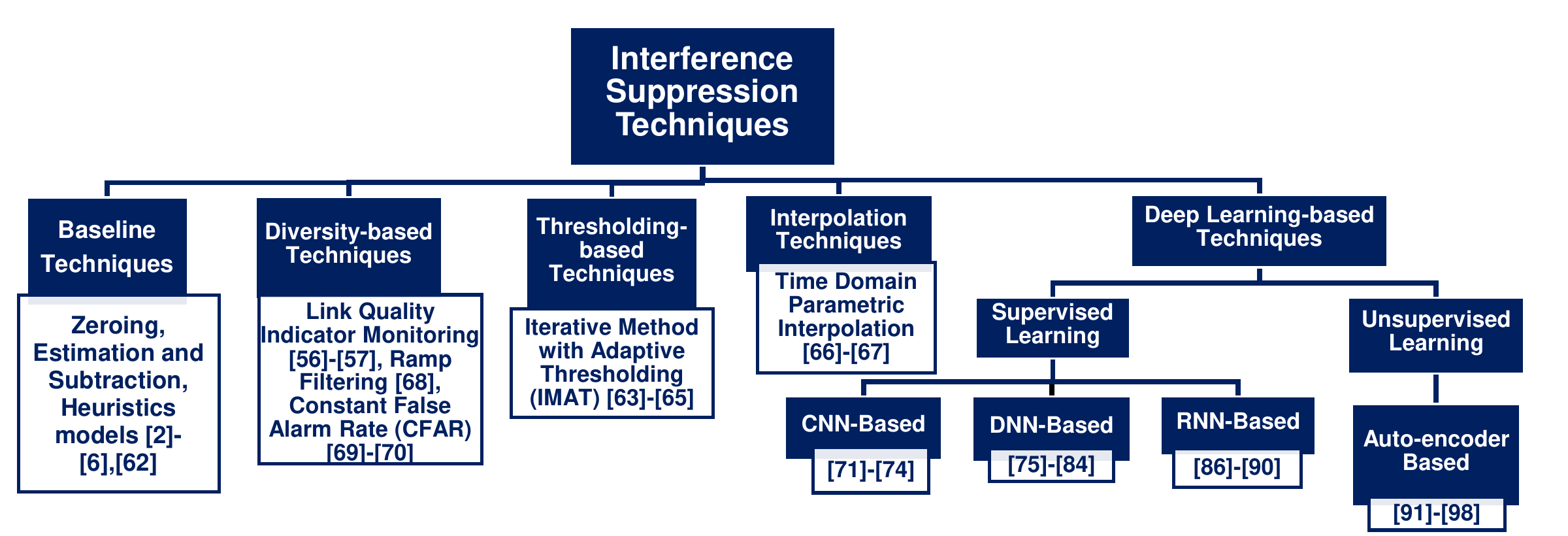}
\caption{Interference Suppression Techniques}
\label{fig:techniques}
\vspace{-0.2in}
\end{figure*}

\subsection{Classification of Interference Suppression Techniques}
\label{conventional}
Broadly speaking, interference suppression can be divided into the following categories (shown in Figure \ref{fig:techniques}): baseline techniques, diversity-based techniques, thresholding-based techniques, interpolation techniques and deep learning-based techniques. Furthermore, researchers have investigated polynomial-based cancellation techniques to cancel SI \cite{korpi2017nonlinear,anttila2014modeling,islam2019comprehensive}. Polynomial-based methods used in full duplex systems typically result in a huge number of parameters which further make their implementation complex. While this survey focuses on deep learning-based techniques, we briefly describe non-deep learning techniques in this section to provide some background information necessary to understand the comparison in Section \ref{sec:comparison}. Fig. \ref{fig:techniques} shows a taxonomy that includes both non-deep learning and deep learning techniques for interference suppression.

Before the adoption of machine learning or deep learning methods to suppress interference, researchers used various techniques to suppress or excise them. Because many low-power wireless networks are power- and computation-constrained, PHY layer information and time diversity of the wireless channel have been proposed to suppress heterogeneous technology interference. For example, Hithnawi \cite{hithnawi2013exploiting} monitored link quality indicator (LQI) \cite{hermans2012lightweight} values of corrupted frames together with conventional threshold mechanisms to detect the presence of interference. When interference is detected, the changes in received signal strength indicator (RSSI) is collected. The RSSI changes are captured by sampling the received signal. The RSSI readings is then sent to the detection algorithm. The corrupted packet is reconstructed from subsequent interfered re-transmission and the buffering of the correctly decoded bytes \cite{hithnawi2013exploiting}. 

While we acknowledge that earlier works have investigated interference avoidance schemes, such as transmit power control \cite{vanheel2007spectral}, dynamic frequency switching \cite{yoon2010adaptive,yun2008channel,hassan2013mitigating}, amongst others, our survey focuses on techniques that suppress interference. These techniques, shown in Fig. \ref{fig:techniques}, include zeroing of interference samples \cite{fischer2016untersuchungen}, iterative method with adaptive thresholding (IMAT) \cite{bechter2017automotive,marvasti2012sparse,alland2019interference}, time domain parametric interpolation \cite{tullsson1997topics,chan1981parameter}, ramp filtering (RFmin) \cite{wagner2018threshold,wang2021cfar,richards2005fundamentals} amongst other approaches. While zeroing out interference components may be the simplest method in suppressing interference, it could reduce SNR or even potentially zero out actual signal \cite{toth2019performance}. The RFmin technique exploits the diversity of range spectrum over ramps \cite{toth2019performance}. In IMAT approaches, the interfered samples are first zeroed  and the corresponding zeroed time-domain gap is then iteratively interpolated  by replacing the gap with the inverse Fourier transform of the actual dominant spectral component \cite{bechter2017automotive,marvasti2012sparse,toth2019performance}. The time domain parametric interpolation approach is similar to the IMAT approach with the difference being the use of a parametric signal model to interpolate the gaps resulting from zeroing out interfered signals \cite{tullsson1997topics}. It is important to point out that the list of non-deep learning techniques for interference suppression mentioned in this section is non-exhaustive, we only highlight a few of those techniques since the main focus of this survey paper is deep learning techniques for interference suppression. Readers who are interested in these other techniques can refer to \cite{kunert2010d1,lu2006survey,le2015interference}. Fig. \ref{fig:techniques} shows popular algorithms for each category.

Deep learning techniques are generally divided into supervised and unsupervised learning\footnote{The major distinction between supervised and unsupervised learning is the use of labeled datasets. The former relies on labeled inputs and outputs while the latter uses machine learning algorithms to discover the structure of the unlabeled input data}. For the supervised learning techniques, CNN-based suppression techniques were discussed in \cite{ristea2020fully,rock2019complex,rock2021resource,fuchs2021complex}, deep neural network (DNN) based techniques were discussed in \cite{xu2018deep,hager2018nonlinear,promsuk2021development,mosavi2016narrowband,shi2012interference,muranov2021deep,balatsoukas2018non,guo2019dsic,shi2019digital,kurzo2018design,zhang2018self,yang2010iterative}, and recurrent neural network (RNN) based techniques were investigated in \cite{mun2018deep, mun2020automotive,wei2016gps,chang1999narrow,xu2008narrowband}. For the unsupervised learning techniques, autoencoder-based suppression techniques were discussed in \cite{lin2020cross,fuchs2020automotive,dorner2017deep,o2017introduction,erpek2018learning,wu2020deep,wu2019adaptive,chen2021dnn}. 

Furthermore, machine learning (ML) based channel state information (CSI) techniques \cite{lynggaard2018using} have also been studied to estimate the required transmit power level necessary to suppress interference. This approach strongly relies on a precise estimation of transmit power needed to suppress interference in a wireless sensor network (WSN) application. Other methods, such as \cite{suzain2020machine,hermans2012lightweight} have utilized lightweight feature extraction and lightweight decision trees to detect and identify interference in Wi-Fi, Bluetooth, and microwave ovens.


Because of the plethora of devices contending to share the wireless spectrum, conventional interference suppression approaches may not be scalable to resolve associated interference issues that may result. Since deep learning approaches have been found to improve classification accuracy for many interference classification or identification schemes\cite{schmidt2017wireless,kim2020classification,pulkkinen2020understanding,grunau2018multi}, researchers have started taking their investigation to the next level and their potentials for interference suppression \cite{rock2019complex,ristea2020fully,mun2018deep, fuchs2020automotive}. We discuss them in detail in section \ref{sec:DL_IM}.

\subsection{Deep Learning Fundamentals}\label{sec:DL}
In the last two decades, we have witnessed tremendous and unparalleled breakthrough in computing hardware technologies. This breakthrough led to affordable and pervasive computing resources. Take, for example, graphics cards which have seen significant growth in computing power. Computers are now capable of doing complex operations in such a short amount of time. Deep learning techniques allow computers to build complex concepts out of simpler concepts \cite{goodfellow2016deep}. Broadly speaking, there are three types of deep learning algorithms: supervised, unsupervised and reinforcement learning. Supervised learning techniques typically learn from labeled data \cite{caruana2006empirical,hastie2009overview} while unsupervised techniques do not require labels to learn from the data \cite{bengio2012deep}. Reinforcement learning \cite{sutton2018reinforcement} on the other hand deals with an intelligent agent that take actions in order to maximize its cumulative reward. Semi-supervised learning \cite{zhu2009introduction,chapelle2009semi,berthelot2019mixmatch}, a technique that combines supervised (labeled data) and unsupervised (unlabeled data) learning has also received a lot of attention in recent times. While earlier works such as Longi et. al. \cite{longi2017semi} utilized a semi-supervised learning technique to detect interference sources from wireless local area networks (WLAN), we note that, to the best of our knowledge, their use in interference suppression has received little to no attention. 

Deep learning techniques have enjoyed significant success in solving challenging problems in several application domains, including image classification, medical bio-informatics, natural language translation amongst others. Motivated by these developments, researchers in wireless communication have embraced deep learning to address the most difficult problems in wireless communications and networking, such as interference identification, classification \cite{schmidt2017wireless,grunau2018multi} and suppression \cite{fuchs2020automotive}, modulation recognition \cite{o2018over}, transmitter classification \cite{morin2019transmitter}, spectrum sensing \cite{lee2017deep}, traffic classification \cite{aceto2019mobile} amongst other challenges. In addition, deep learning techniques can be very useful in managing coexistence issues arising from the deployment of fifth generation (5G) New Radio (NR) and sixth generation (6G) technologies with regards to scalability, generalization and the ability to learn from the data rather than reliance on domain expertise. Therefore, using deep learning to estimate, predict and suppress time-varying interference strengthens interference resilience of the PHY layer. 

Earlier applications of deep learning to wireless communications have investigated DL-based end-to-end systems for multiple user, multiple antenna systems that mutually optimize the transmitter and the receiver so as to improve performance compared to traditional non-DL or non-ML approaches \cite{o2017physical,erpek2018learning}. Deep learning techniques have also been used for channel modeling and estimation. For instance, generative adversarial networks (GANs)\footnote{GANs learn deep representations by training a pair of networks in competition with each other. It consists of a generator and a discriminator network in which the generator aims to trick the discriminator by generating fake samples. The discriminator on the other hand receives both the fake samples and the real samples and tries to tell them apart. Both generator and discriminator are usually implemented with neural networks \cite{goodfellow2014generative}. Hence, one neural network is pitted against another neural network with each improving its capabilities over time.} were used by \cite{o2018physical,ye2018channel} to approximate stochastic channel response from the raw data for real time channel modeling. Furthermore, DL-based OFDM receivers have been used to learn the channel between a transmitter and a receiver and also decode the received signal \cite{ye2017power}. DL-based belief propagation algorithm that cascades CNN to a belief propagation network was utilized by Liang \textit{et al.} \cite{liang2018iterative} to accurately estimate noise. While these techniques are not interference suppression techniques, they may provide useful information for subsequent interference suppression processing. 

\subsubsection{Convolutional Neural Network}
CNNs are supervised deep learning techniques that learn the relationship between input data and their corresponding labels \cite{albawi2017understanding}. For instance, in \cite{akeret2017radio}, the authors proposed an approach to suppress radio frequency interference (RFI) in radio frequency data using the U-Net. A U-Net is a CNN algorithm that is used to classify clean signal and RFI signatures in 2D time-ordered data acquired from a radio telescope. While this work did focus on the use of a deep learning algorithm to classify interference, the technique used was only validated for cosmology applications that deals with satellite systems, air traffic communications, radio telescope, etc. Just like RFI suppression research, radar interference suppression is also another well-studied research area \cite{mun2018deep,ristea2020fully,fuchs2020automotive,fan2019interference,rock2019complex,rock2020quantized}. Using CNNs to suppress interference is still an emerging area of research. This is because there are many factors to the successful adoption of CNNs to interference suppression. One such factor is the availability of benchmark datasets.

\begin{figure*}[!bt]
\centering
\includegraphics[scale=0.425]{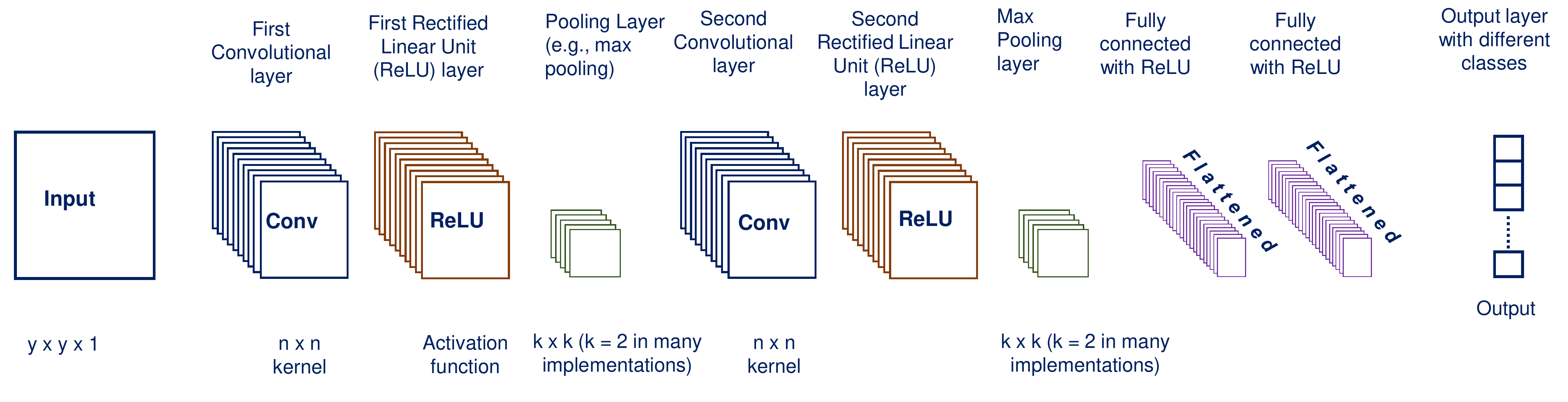}
\caption{A generic CNN architecture with convolution layers, ReLU activation function, pooling layer interlaced with each other and two fully connected layers}
\label{fig:cnn}
\vspace{-0.2in}
\end{figure*}

Unlike in other deep learning application domains where there is a plethora of publicly available benchmark datasets, only a handful of datasets are available for interference suppression research. In addition, it is quite difficult to create a dataset that can be used for training and testing different models to address different problems in various wireless communications/networking applications. This is because the wireless channel is stochastic in nature and there can be different realizations at different points in time. More so, wireless channels are highly dynamic, it is very likely the training data generated in one scenario could not be applicable to other scenarios. Hence, it is not surprising that there are no publicly available benchmark datasets for the vast majority of wireless research problems. Researchers working on these problems need to generate their own datasets, which is very resource-intensive and time consuming. However, datasets that capture unique environments are still valuable in training deep learning algorithms despite the fact that they are not representative of all channel environments. As noted earlier, SDRs can be leveraged or collecting data in multiple channel environments.

Fig. \ref{fig:cnn}  shows a generic CNN architecture with convolution layers, rectified linear unit (ReLU) activation layers and pooling layers interlaced with one another before the output is then sent to two fully connected layers. This generic architecture is similar to the one described by \cite{o2015introduction}. The input data could be images as seen in the MNIST data with size $28 \times 28 \times 1$ \cite{kussul2004improved}. The input image is then passed through the convolution layer that uses convolution filters or kernels to run over the input array and compute a dot product in order to extract different features from the input array. The kernel could be slid in increment of one or more cells, this is usually referred to as the stride. In order to introduce non-linearity to the model, activation functions are used to learn complex mappings between the input data and the response variables. Examples of activation functions include ReLU, sigmoid, tanh, leaky ReLU, etc. The ReLU activation function is the most common of them because it is a piece-wise linear function defined as follows

\begin{equation}
    ReLU(x) = max(0,x).
\label{relu}    
\end{equation} 

From equation \ref{relu}, if $x > 0$, the function is positive and if $x < 0$, the output will be zero. The pooling layer decreases the size of the features that were convolved so as to reduce overfitting. The fully connected layer is very useful in capturing non-linear combinations of the high-level features. There could be more than one fully-connected layers. Depending on the application, the last fully-connected layer's output dimension is based on the number of classes. CNNs are capable of learning over both time and frequency domain, hence, their usefulness in modelling devices with output that is not restricted to frequency outputs alone \cite{pulkkinen2020understanding}. Furthermore, CNNs are very useful when the data captures non-linearities in the time domain specifically for identification or classification tasks \cite{merchant2018deep}. 

Prior studies have demonstrated that CNN can automatically extract device-specific features by analyzing a huge amount of training data to improve the performance. For instance, authors in \cite{merchant2018deep} presented deep learning techniques useful in fingerprinting  several IEEE 802.15.4 devices. Further, another work \cite{youssef2017machine} investigated different the combination of machine learning algorithms with wavelet transforms to uniquely identify and classify disparate wireless transmitters. On a different note, DL-based modulation recognition solutions have also attracted much attention in recent years. For example, \cite{o2018over, peng2018modulation} leveraged CNN algorithms to recognize modulation schemes based on sequences of constellation points. Several other papers \cite{wang2019data,karra2017modulation,xie2019deep} studied the application of deep learning to modulation recognition problems.

Some authors investigated interference classification using deep neural networks. For instance, \cite{yu2020interference} developed a technique for classifying interference at the PHY layer. The interference classified include single tone, chirp, filtered noise and unknown modulated signal. They extracted cyclic spectrum and spectrum information from the received signal. 
While their overall finding identified which spectrum information feature to extract, they did not consider ways to suppress interference sources.

Grunau \textit{et al.} \cite{grunau2018multi} proposed a wireless interference identification that classified multiple Bluetooth, Wi-Fi and IEEE 802.15.4 interfering signals in the presence of the desired signal using a CNN. In their work, they generated a \textit{multi-label} dataset that consists of time- and frequency-limited sensing snapshots that combines the desired signal with multiple interference signals. While their work classified the interference sources with high accuracy, it can be argued that given the apparent dissimilarities between the three technologies, the task may have been trivial for a DL classifier. Furthermore, there is no discussion on how to suppress the interference from the different technologies. For interference suppression, however, authors such as \cite{rock2019complex,ristea2020fully} have shown that CNNs can learn to suppress interfering signals. We discuss these techniques in section \ref{cnn_suppression}.  



\subsubsection{Autoencoders}
Autoencoders (AE) have been studied as a technique to suppress interference in applications such as radars and Wi-Fi. AE are neural networks that aim to learn a compressed latent-space representation of the input data, that can be successfully reconstructed into the original data \cite{doersch2016tutorial}. The latent-space representation can usually encode useful information about the interfering signal that can then be suppressed at the decoder. They are composed of the encoder which compresses the input into a latent-space representation and can be generally represented by the function: $h = f(x)$ while the decoder reconstructs the input from the latent representation. The decoder can be represented by the function: $x' = g(h)$. For an autoencoder, the main task is to ensure that $x'$ is as close as possible to the original input, $x$. Fig. \ref{fig:ae} shows a block diagram of a generic autoencoder with encoder and decoder networks.

The latent space representation, $h$, has very useful properties since it learns very important features in the input data. At a high level, we can picture the autoencoder as a way of copying the input to the output. However, when the copying process is constrained, such as, limiting $h$ to have a dimension smaller than $x$, the autoencoder is compelled to learn the most important features of the training data \cite{hubens_2018}. On the other hand, if the autoencoder has too much capacity, it may end up not learning vital information about the distribution of the input data \cite{hubens_2018}. The imposed constraint is analogous to intentionally introducing a restriction or bottleneck in the network which results in a compressed useful representation of the original input. The lack of dependence between the input features could make reconstruction difficult. On the other hand, if there is some correlation between the input features, the imposed bottleneck can be leveraged to learn the structure of the input \cite{morawski2021anomaly}. Without the bottleneck or constraint, the network simply memorizes the input values and passes it along through the network to the output. The loss function is usually either mean square error (MSE) or reconstruction loss which is the cross-entropy between the input and the output. The network is penalized for creating outputs different from input. To prevent the autoencoder from simply memorizing the input, it is tuned in such a way that it is sufficiently sensitive to the input such that it can precisely build a reconstruction while also adequately insensitive enough to not overfit the training data \cite{jordan_2018}. One potential application of autoencoders for interference suppression is that $h$ can also be pruned to suppress or reject interference.

\begin{figure*}[!tb]
\centering
\includegraphics[scale=0.65]{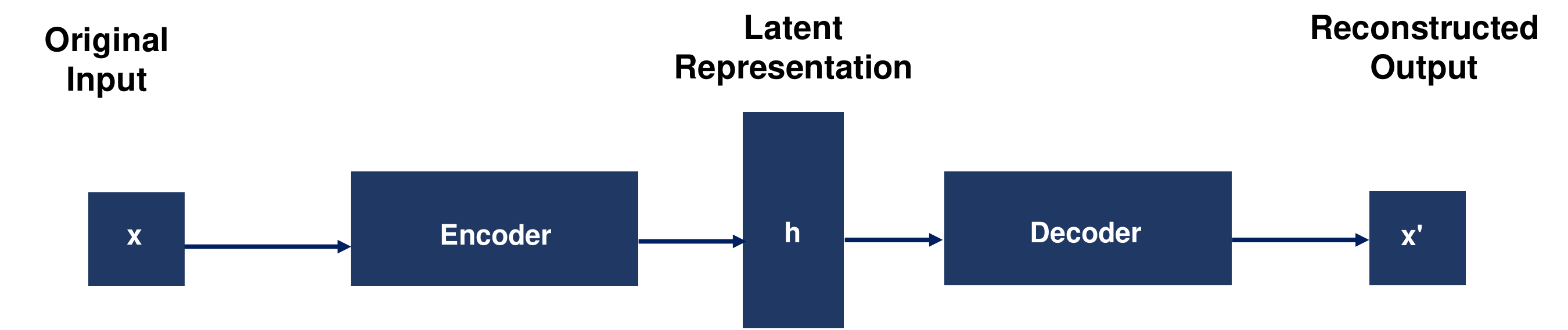}
\caption{A Generic Autoencoder Architecture}
\label{fig:ae}
\vspace{-0.2in}
\end{figure*}

Some general applications of autoencoders include data denoising\cite{meng2017research}, dimensionality reduction\cite{wang2016auto}, anomaly detection\cite{zhou2017anomaly}, image in-painting\cite{shcherbakov2014image}, information retrieval\cite{feng2014cross}, etc. In data denoising applications, an AE learns a generalizable encoding and decoding function, thus, corrupting the input data with noise may not significantly impact accurate reconstruction. On the other hand, for dimensionality reduction applications, when compared to principal component analysis (PCA) or other dimensionality reduction techniques, autoencoders are able to easily learn data projections that are more useful \cite{hubens_2018}. This helps to map from high feature space to low feature space. As we discuss in section \ref{ae}, these unique features of autoencoders made them very crucial to detecting and suppressing interference. In section \ref{sec:dataset}, we discuss publicly available interference datasets.

\section{Publicly Available Interference Dataset}
\label{sec:dataset}
One of the key requirements for using deep learning to solve any problem is the access to benchmark datasets. Take for instance, in the image processing/computer vision and the medical bioinformatics research areas where deep learning techniques are very pervasive, there are a plethora of benchmark  datasets available for use by their corresponding research communities. It is important to also point out that utilizing simulated and synthetic dataset is frowned upon by deep learning experts in fields such as computer vision and image processing. This is because the performance of such algorithms on these dataset types often skew their expectation for real world applications. For example, data collected in a simulated environment with AWGN channel and fading differ significantly from data collected over-the-air (OTA). Furthermore, access to benchmark datasets helps to further promote cutting edge research as well as innovation in the field. For instance, in image processing domain, the modified National Institute of Standards and Technology (MNIST) database consists of $70k$ images for both training and testing \cite{kussul2004improved} while there is also an extended dataset, EMNIST \cite{cohen2017emnist} that has $280k$ images of handwritten digits and characters. Despite the paucity of such benchmark datasets in the interference suppression research, there are still a few authors and useful community resources where interference datasets can be accessed. In this section we provide the readers, who are interested in deep learning-based interference classification or suppression, high level information on these resources. Using these datasets and with appropriate scaling and sample rate conversion, synthetic datasets can be created using real data.

\subsection{Radar Interference Dataset}
Ristea \textit{et al.} \cite{ristea2020fully} automatically generated $48k$ samples of an automotive radar dataset with only one interference source. While this may seem insufficient, it is indeed a good contribution to the largely unavailable public generated databases for interference suppression. They fixed parameters such as bandwidth, sampling frequency, center frequency and sweep time while other parameters such as SNR, SIR, target amplitude, distance, and phase were selected using a uniform distribution between a lower and upper bound. Details of these values can be found in \cite{ristea2020fully}. Their data samples contain three categories of data such as time domain signal with no interference, time domain signal with interference and a label vector with complex amplitude values in target location. The dataset is available for download in \cite{ristea2020dataset}. Radar signals are especially important datasets because of the extensive spectrum sharing interference between radar and communication signals, and the dynamic behavior of radar makes it difficult to simulate. 

\subsection{Radio Frequency Interference Dataset}
The radio frequency interference (RFI) dataset was generated by authors of paper \cite{ujan2020efficient}. The dataset was created by combining the signal of interest (SOI) with three popular jammers: continuous-wave interference (CWI), multi-continuous-wave interference (MCWI) and chirp interference (CI). In addition, the dataset was generated under different signal to noise ratios (SNR). The desired signal, a real-time video stream, was modulated using GNU radio and then transmitted via a Universal Software Radio Peripheral (USRP-N210). The authors utilized a combiner to combine the respective jamming signal with the SOI. The receiver used was a MegaBee modem. Details of their dataset generation can be found in \cite{ujan2020efficient}. Fig. \ref{fig:rfi} shows a scalogram plot\footnote{The scalogram represents the squared magnitude of continuous wavelet transform (CWT), a type of wavelet transform, which is used for non-stationary and transient signal analysis. The CWT is a non-numerical signal transformation function that continuously varies translation and scale parameter of the wavelets. It essentially provides an detailed or over-complete representation of a signal. Authors in \cite{slavivc2003damping} noted that CWT is resistant to noise in the signal .} of the RFI dataset. It is clear from the scalogram plot that distinct features of each class of interference can be easily learned by a deep learning algorithm.

\begin{figure*}[!tb]
\centering
\includegraphics[scale=0.375]{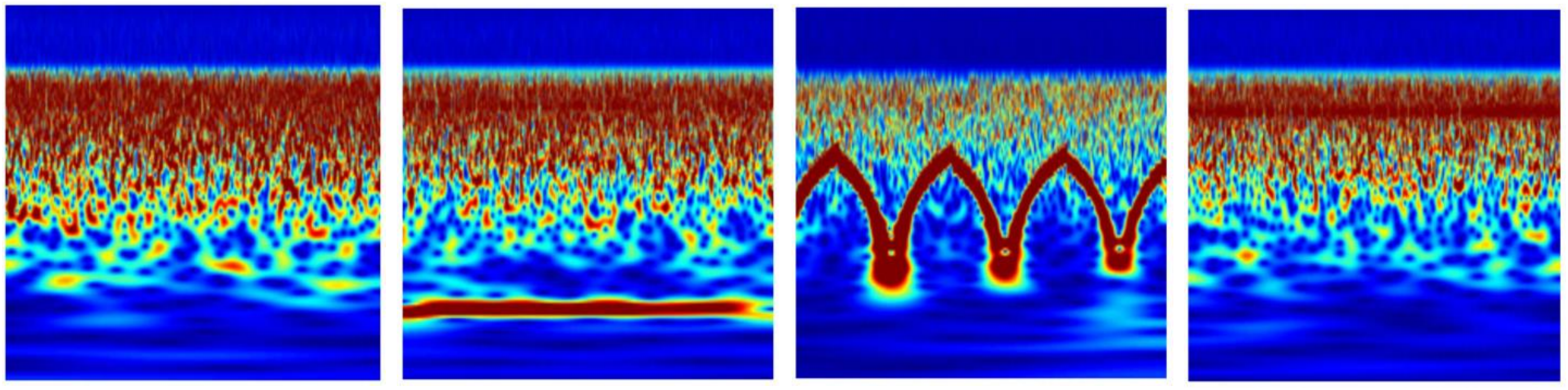}
\caption{A scalogram plot of the RFI dataset from \cite{ujan2020efficient}. (\textit{Left: The original signal of interest with no interference; Center Left: The signal of interest with multi-continuous wave interference;  Center Right: The signal of interest with chirp interference; Right: The signal of interest with continuous wave interference}). The figure was taken from paper \cite{ujan2020efficient} titled: \textit{An Efficient Radio Frequency Interference (RFI) Recognition and Characterization Using End-to-End Transfer Learning}; published as an open access article in the Journal of Applied Sciences.}
\label{fig:rfi}
\end{figure*}

\subsection{CRAWDAD Dataset}
The Community Resource for Archiving Wireless Data at Dartmouth (CRAWDAD) was established to bridge the gap in collecting data captured from production wireless networks so as to understand how real users and devices use networks under real conditions \cite{crawdad}. At the time of writing, there are 125 datasets for different wireless communication applications and more than $14k$ users from more than $120$ countries \cite{crawdad}. The dataset, contributed by Schmidt \textit{et al.}  \cite{schmidt2017wireless}, contains traces of IEEE 802.11b/g, IEEE 802.15.4 and Bluetooth packet transmissions with varying SNRs in the baseband \cite{owl-interference-20190212}. The authors also added frequency offsets in the baseband \cite{grunau2018multi}. In total there are fifteen different classes with ten of those classes being IEEE 802.15.1 devices, three being IEEE 802.11 and the remaining two being IEEE 802.15.4 devices. There are two other interference datasets on the CRAWDAD database that are targeted towards the measurement of HTTP requests over 802.11 protocol in a dense wireless classroom \cite{kyutech-interference-20150616} and a trace-set that includes received signal strength indicator (RSSI) for frames that were correctly received at receiver nodes when noise was injected \cite{rutgers-noise-20070420}. For more on the CRAWDAD datasets, readers are directed to \cite{crawdad}.

\subsection{Wireless Interference Signal Dataset}
Authors in \cite{grimaldi2018real} collected experimental data in four different environments with one of those environments having no detectable interference, another with controlled interference and the other two environments with uncontrolled interference from real devices. Interference sources for their experiments include Bluetooth and Bluetooth Low Energy (IEEE 802.15.1 and IEEE 802.15.1 BLE respectively), ZigBee (IEEE 802.15.4) and WLAN (IEEE 802.11). In total their dataset consists of more than $110k$ interference bursts of known origin. One key contribution of their work is the heterogeneous nature of the interference scenarios they considered which makes for a robust dataset that can be used by the wireless community at large.

Table \ref{table:Dataset Comparison} shows a high level comparison between the datasets mentioned in this paper. 

\begin{table*}[!bt]
\caption{Comparison Between Interference Datasets}
\centering
\footnotesize
\begin{tabular}{|p{4.5cm}|p{5cm}|p{5.5cm}|}
\hline
\textbf{Dataset} & \textbf{Interference type}   & \textbf{Signal Types}\\
\hline
\hline
Radar Interference Dataset \cite{ristea2020dataset} & Homogeneous technology interference & $3$ classes including: \begin{itemize}\item Time domain signal with interference \item Time domain without interference \item Amplitude information for targets \end{itemize}\\
\hline
Radio Frequency Interference Dataset \cite{ujan2020efficient} & Heterogeneous technology interference & $4$ classes including: \begin{itemize}
    \item Signal of Interest (SOI) \item Continuous-wave interference (CWI) \item Multi-continuous-wave  interference (MCWI) \item Chirp interference (CI)
\end{itemize} \\
\hline
CRAWDAD Dataset \cite{schmidt2017wireless} & Homogeneous and heterogeneous technology interference & $15$ classes including \begin{itemize}
    \item Bluetooth ($10$) \item WLAN ($3$) \item ZigBee ($2$) 
\end{itemize}\\
 \hline
Wireless Signal Interference Dataset \cite{grimaldi2018real}  & Heterogeneous technology interference & $4$ classes including: \begin{itemize}
    \item Bluetooth \item Bluetooth Low Energy (BLE) \item ZigBee \item WLAN
\end{itemize}\\
 \hline
\end{tabular}
\label{table:Dataset Comparison}
\end{table*}

\section{Deep Learning-Based Interference Suppression Techniques}\label{sec:DL_IM}
In this section, we survey the most popular DL-based interference suppression techniques shown in Figure \ref{fig:techniques}. The most common deep learning techniques discussed in this section include CNN, autoencoders, RNN and deep neural networks (DNNs). While there are other deep learning techniques in general, our research shows that these four techniques are the most common in the interference suppression literature.

\subsection{Contributions for Interference Suppression Using CNN} \label{cnn_suppression}
In this section, we investigate CNN-based techniques for interference suppression. CNNs, an example of a model-free method, have shown impressive performance in many other domains (e.g. radio fingerprinting \cite{merchant2018deep}, image classification \cite{cohen2017emnist,ciregan2012multi}, etc). CNNs can utilize only raw IQ data or other channel-based information to learn the unique features in interfering signals \cite{xu2020open, kulin2018end, kim2018identifying}.  Typically, CNN based techniques include layers of convolution, pooling ReLU and batch normalization operations as shown in Fig. \ref{fig:cnn_rock}. CNNs, unlike feature-based machine learning (ML) techniques do not require that a domain expert handcraft the features. Instead, the algorithm through the use of convolutional layers and a feature map, learns salient features necessary for suppressing interference. CNNs are capable of learning over both time and frequency domain, hence, they are useful in modeling devices with output that is not restricted to frequency domains alone \cite{pulkkinen2020understanding}. 
While earlier works \cite{kim2020classification,schmidt2017wireless,pulkkinen2020understanding,grunau2018multi} have investigated interference classification using CNN, several recent works have shown that CNNs can also go beyond classifying interference to actually suppressing them. For instance, Rock et. al. \cite{rock2019complex} utilized CNNs to find structured information in the input data to suppress automotive radar interference. One key application of their CNN model is to de-noise and suppress interference using a dataset that consists of clean data and data with interference. They used metrics such as signal to interference ratio (SIR) and signal to noise ratio (SNR) to scale the interference and noise powers in relation to the actual object signal power. In their CNN model (shown in Fig. \ref{fig:cnn_rock}), the first layer consists of a convolution layer and a ReLU activation function while the next four layers consist of batch normalization\footnote{Batch normalization is usually used in deep learning training to reduce the number of epochs and stabilize the learning process by standardizing the inputs to a layer for every mini-batch \cite{ioffe2015batch}.}, convolution and ReLU activation function operations, the last layer (before the output layer) consists of a batch normalization and convolution operations. Convolutional layers in CNNs often learn unique artefacts in the input data that help in identifying interfering signals in the data. Their results show that the CNN based suppression approach outperform non-deep learning techniques such as zeroing, IMAT and RFmin. For instance their CNN based suppression model outperformed zeroing, ramp filtering and IMAT techniques by more than 40\% using the signal to interference plus noise ratio (SINR) metric \cite{rock2019complex}. Despite these gains, the CNN-based suppression techniques suffered a relatively high error vector magnitude (EVM) performance when compared to the conventional suppression approaches such as zeroing, IMAT and RFmin. For radar applications, such as the one considered in the authors' study, a high EVM may result in distortions in object peak values \cite{rock2019complex}. It is important to point out that the study done by Rock et. al. \cite{rock2019complex} utilized a simulation framework. This may imply that some of parts of their results may not hold for over-the-air (OTA) applications. This is because real world OTA applications may be subject to more channel uncertainties such as fading, noise, mobility and other factors. It is fair to say that one might expect different results for QAM communication signals with information carried in the signal phase.

\begin{figure*}[!bt]
\centering
\includegraphics[scale=0.35]{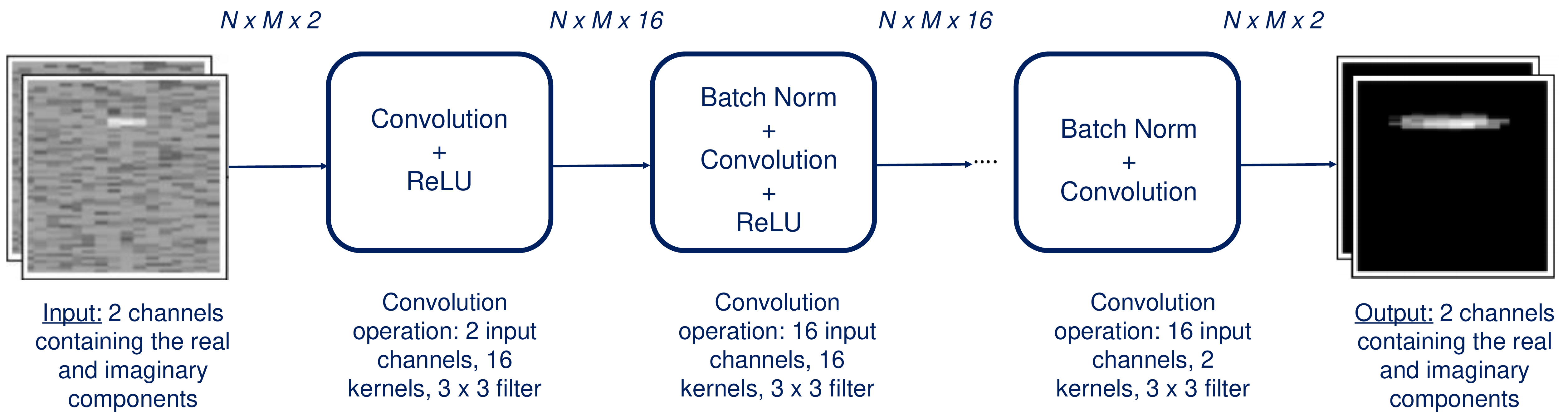}
\caption{CNN Architecture proposed by Rock \textit{et al.} \cite{rock2019complex} for radar signal denoising. The figure was reproduced from paper \cite{rock2019complex} titled: \textit{Complex Signal Denoising and Interference Mitigation for Automotive Radar Using Convolutional Neural Networks}; published at the 22nd IEEE International Conference on Information Fusion (FUSION).}
\label{fig:cnn_rock}
\vspace{-0.2in}
\end{figure*}

Similar to \cite{rock2019complex}, Ristea \textit{et al.} \cite{ristea2020fully} utilized a fully convolutional neural network, another CNN approach, to learn the patterns in the input data using a small amount of learnable parameters. They proposed two deep learning architectures that outperformed conventional zeroing techniques of interference rejection \cite{fischer2016untersuchungen} by taking spectrograms of input signals (with $1024$ samples) as input and then arriving at clean signals at the output. The input signals consist of noise and uncorrelated interference which is learned using the fully convolutional neural network (FCN). To obtain the beat signal, they first convert the time domain signal into a spectrogram by using short time Fourier transform (STFT). To get the ground truth, the authors perform an FFT on the time domain labels and consequently trained the network to map the STFT input to the FFT output. Generally, in the wireless communication community, many researchers utilize the I/Q samples to train the deep learning architecture. A good lesson from this paper stems from the row dimension of the I/Q samples having a value of $2$, the filters used in the convolutional layer may be too small and hence leading to overfitting or further complicating the CNN algorithm. This is because using small sized filters tend to make the model less generalizable, and thus, learn specific features in the training data. 
In another work, Rock \textit{et al.} \cite{rock2021resource} investigated CNN-based quantization techniques to mitigate or de-noise interfering radar signals. Specifically, their work focused on memory and computational requirements associated with using deep learning models. The major motivation for this is that high memory requirements could limit the energy efficiency and the time it takes to execute operations when an embedded hardware is used \cite{rock2021resource}. To account for this, the authors explore quantization capabilities of different CNN architectures. Their results showed that the CNN-based mitigation models performed better than the three traditional interference mitigation methods considered (zeroing, IMAT and ramp filtering) and are also resilient against different interference levels and patterns. They also considered interference signals from real-sensor measurements that was combined with an object signal from an inner-city measurement campaign, their result showed an SNR improvement of 7dB after CNN-based mitigation \cite{rock2021resource}.

Further, Fuchs \textit{et al.} \cite{fuchs2021complex} proposed a complex valued CNN (CVCNN) to suppress mutual interference in automotive radar systems. The main challenge the authors sought to address was the problem of reconstructing the phase of clean signals from the real and imaginary parts of the complex spectra. This is because many object detection algorithms typically rely on only the magnitude spectrum component of the interference mitigated maps, and thus may not include important phase information. While including the phase information adds more complexity to the algorithm, complex-valued analysis can also restrict the degrees of freedom \cite{fuchs2021complex}.  In \cite{fuchs2021complex}, the authors found that such a restriction can leverage the physical characteristics of the data to enforce signal transformations thereby reducing the complexity. Their CVCNN model, thus, uses complex-valued analysis for all the operations. This implies that the interconnection between real and imaginary parts in the model architecture are taken into consideration. This also implies that the algorithm does complex convolution, complex batch normalization and complex ReLU activation function. Their results showed that the CVCNN method outperformed the real-valued CNN, zeroing, RFmin and IMAT techniques in terms of F1-score, EVM and peak phase MSE (PPMSE)\footnote{PPMSE is the measure of the average squared difference between the angle of the clean range-Doppler signal vector and the predicted range-Doppler signal of all detected object peaks \cite{fuchs2021complex}.}.

In practice, a CNN model should be able to generalize well so that it can appropriately classify samples from the test data. To solve this, many researchers use preprocessing techniques such as spectrograms \cite{ristea2020fully} or continuous wavelet transforms (CWTs) \cite{oyedare2019estimating,youssef2018machine} to further increase the dimension of the input data along the time-frequency axis. It is very important to preprocess time domain signals to produce a 2D image representation so that CNNs can better learn salient features in the input data. In \cite{ristea2020fully}, the authors used a spectrogram which led to an input size of $154 \times 2048$. Their results showed that the FCN algorithm moderately outperformed the conventional suppression method such as zeroing using the area under the receiver operating characteristics curve (AUC) metric. We note that their technique improves the SNR and AUC scores of the actual signal and not just the spectral representation. This goes to show that combining traditional signal preprocessing techniques with deep learning can significantly improve interference suppression performance while reducing overfitting.

\begin{figure}[!bt]
\centering
\includegraphics[scale=0.32]{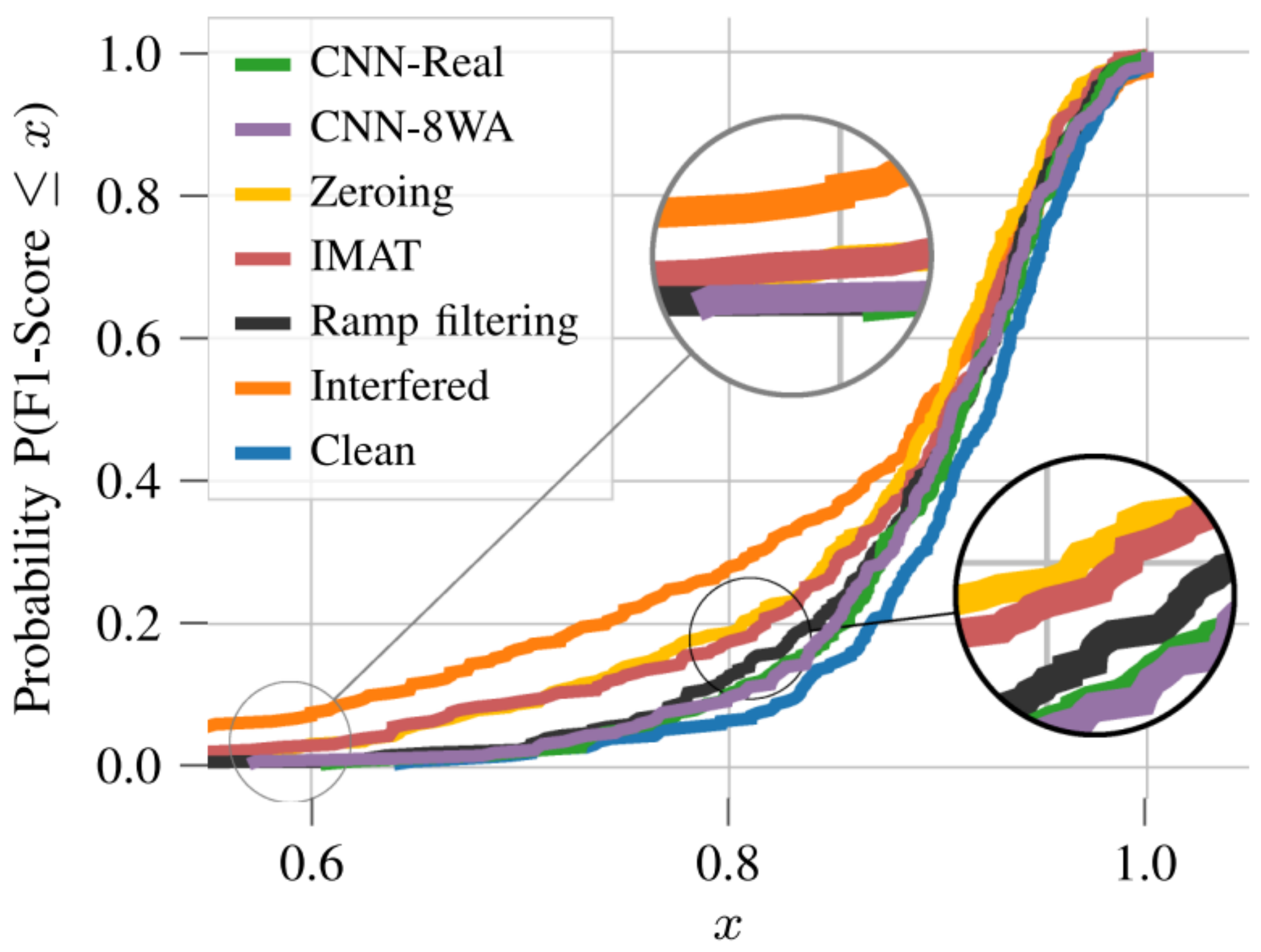}
\caption{A CDF comparison of the sample-wise F1-Score between two CNN models, zeroing, IMAT and Ramp filtering interference mitigation techniques (\textit{where $x$ is a real-valued number}). The figure was taken from paper \cite{rock2021resource} titled: \textit{Resource-Efficient Deep Neural Networks for Automotive Radar Interference Mitigation}; published in IEEE Journal of Selected Topics in Signal Processing, vol. 15.}
\label{fig:cnn_example}
\end{figure}

It is not surprising to see that the research community is slowly adopting the idea of preprocessing input signal dataset since it provides an opportunity to explore the best of both worlds (i.e. signal processing and deep learning applications). More so, preprocessing techniques (such as continuous wavelet transform (CWT)) can be very useful in learning attributes unique to the transceivers, such as non-linearity in the power amplifier, oscillator imperfections, amongst other unique attributes. In over-the-air transmissions, both wireless interference and the stochasticity of the wireless channel can result in the attenuation of unique attributes in the transmitted signals \cite{polak2011identifying}. To magnify these attributes, raw IQ signals can be pre-processed using signal processing transforms similar to CWT such as discrete wavelet transform (DWT), STFT, Fourier transforms amongst others. Since Fourier transforms does not mark the start and the end of the sine wave, the STFT analyzes several smaller section of the signal. In DWT applications, a sparse representation of the signal is provided. CWT is very useful in enabling a clear representation of many frequency components of a signal as a function of time. The CWT introduces spatial correlations that may not be easily observable in the original signals. The unique features of each transmitter is highlighted in a 2D representation, thus improving the performance of the CNN classifier \cite{oyedare2019estimating}. As a result of this, CWT is better choice for signal or time series analysis because of its fine grained resolution. Such a resolution is obtained by decomposing the signal into wavelets using basis functions (scaled and shifted versions of the mother wavelet \cite{youssef2018machine,polikar1996wavelet}). Finally, a two-dimensional matrix where the rows correspond to the scales and the column corresponding to the length of the signal is then generated. It has been shown that combining signal preprocessing techniques with CNN can significantly improve performance \cite{youssef2018machine}.


\subsection{Contributions for Interference Suppression Using RNN}
Some other works including \cite{mun2018deep} have used recurrent neural networks (RNN) with gated recurrent unit (GRU) to remove automotive radar interference and to simultaneously reconstruct the transmit signal. Unlike many feed-forward neural networks where the collected samples are independent of each other, RNN leverage the dependencies. RNNs capture the dynamics of sequences via cycles in the network of nodes  \cite{lipton2015critical}. Since the received samples are collected sequentially in time domain, RNNs can indeed be used since previous studies show that they are very useful for sequential data. Similar to the deep learning methods discussed, their technique performed better than time domain thresholding described in \cite{watanabe2007interference} and the threshold-free frequency domain interference cancellation method that attenuates the magnitude response of the recorded chirps described in \cite{wagner2018threshold}.

From the foregoing, we observe that Mun \textit{et al.} \cite{mun2018deep} successfully processed sequential data using an RNN model with a gated recurrent unit (GRU). However, they were not able to reconstruct the original signal in the face of interference. To account for the inability of their RNN model to reconstruct the original signal, Mun \textit{\textit{}} \cite{mun2020automotive} proposed an RNN with self attention to mitigate OFDM-based and frequency modulated continuous waveform (FMCW) radar interference. To restore the original signal perfectly, the RNN model learned the information on the surrounding signals and then applied an attention mechanism \cite{vaswani2017attention} to clearly detect the range and velocity of the target signal using an FFT. Specifically, the authors leveraged the attention blocks to capture sequential relationship between the time steps since the interference partially occurred in the entire data. Fig. \ref{fig:rnn_example} shows 2D FFT plot of the received signal with radar interference and the same signal after the radar interference has been removed. This plot shows the power of RNNs in suppressing interference in radar applications. It is important to note that their technique was tested on only simulated data and results of the real world equivalent is not provided.

On a different note, Mao \textit{et al.} \cite{wei2016gps} investigated an interference suppression technique for global positioning system (GPS) using a derivative-free Kalman Filter-based RNN. An adaptive unscented Kalman Filter-based RNN (UKF-RNN) was used to suppress continuous wave interference (CWI), multi-tone CWI, swept CWI and pulsed CWI. While researchers have shown that RNNs are better at modeling non-linear time series systems than signal processing methods \cite{connor1994recurrent}, some interfering signals may be non-stationary, hence, a model that captures this phenomenon is even more useful for suppressing non-stationary interference. In general, results in \cite{wei2016gps} show that their UKF-RNN algorithm increased SNR improvement while also attenuating the mean square prediction error (MSPE) of the received signals in the presence of interference when compared to other conventional methods. With regards to non-stationary interfering signals, the authors' UKF-RNN algorithm offered SNR improvements between 3.34 dB to 8.30 dB more over other methods \cite{wei2016gps}.  

\begin{figure*}[!bt]
\centering
\includegraphics[scale=0.4]{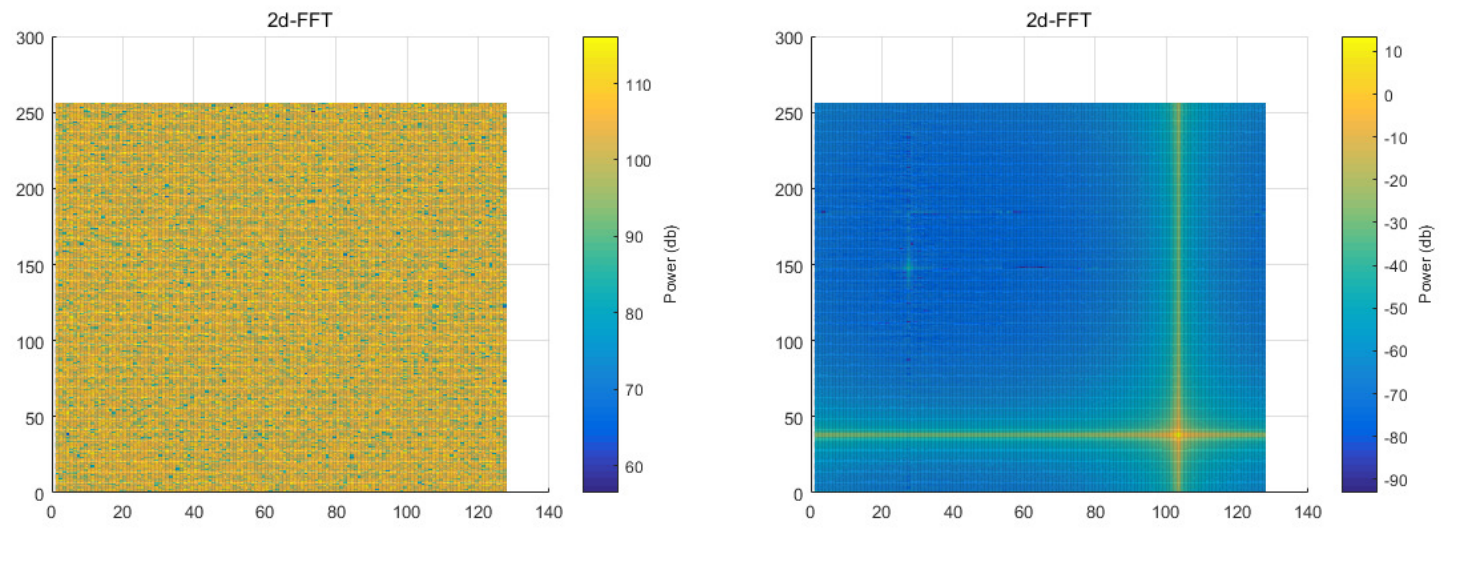}
\caption{A plot showing the 2D FFT result of an RNN-based Interference Mitigation Approach \textit{Left: With radar interference; Right: After radar interference has been removed with RNN.} The figure was taken from paper \cite{mun2020automotive} titled: \textit{Automotive Radar Signal Interference Mitigation Using RNN with Self Attention}; published in IEEE International Conference on Acoustics, Speech and Signal Processing (ICASSP).}
\label{fig:rnn_example}
\vspace{-0.2in}
\end{figure*}

\subsection{Contributions for Interference Suppression Using Autoencoders}
\label{ae}
Another important deep learning technique utilized for interference suppression is the deep autoencoder. Autoencoders are very unique in interference suppression because the encoder network can learn unique interfering components of the input signal (with interference) in the latent space representation. Authors in \cite{fuchs2020automotive} investigated an CNN-based autoencoder technique that was used to suppress automotive radar interference. Similar to \cite{rock2019complex}, their approach involves using a denoising mechanism to suppress interference. They first trained corrupted input signals with their clean non-interfered version. Using deep learning models typically require access to a huge amount of quality data. Their approach ensured that they collected a large amount of range-Doppler images. 

Similarly, Lin et. al. \cite{lin2020cross} proposed a method that used denoising autoencoders to learn patterns of interference with \textit{unknown} structures to suppress heterogeneous technology interference. The main idea behind this approach is to allow the algorithm to learn the distribution of the input data and thus consequently learn the distribution of the interference component so that it can be suppressed. It is important to note that in \cite{lin2020cross}, the authors did not assume any form of coordination or synchronization between the transmitter, receiver and interferer. The authors utilized a generative adversarial network (GAN) to generate synthetic training samples and an autoencoder (denoising) to recover a clean Wi-Fi signal. The synthetic training samples were generated as a means of avoiding the steep cost of collecting a huge amount of data in terms of time and resources. Hence, the GAN was used to synthetically create corrupted Wi-Fi signals that the authors used for training their AE-based suppression model. GAN is basically a network that involves a generator and a discriminator that play a zero-sum game where the generator generates data that is as close as possible to the real data and the discriminator determines whether the generated data  is real or fake \cite{goodfellow2014generative}. Specifically,  their experimental set up included a USRP N200 SDR (within interfering range of a microwave oven) that was deployed as a receiver to collect the signal transmitted when the microwave oven is turned on while Wi-Fi signals are being transmitted. One key lesson learned from their work is that denoising autoencoders perform well when the interfering power is larger than the amplitude of the desired signal (Wi-Fi, in their scenario). Another key benefit of this approach is that it does not incur any signalling overhead since the algorithm itself is passive \cite{lin2020cross}.

Further, Chen \textit{et al.} \cite{chen2021dnn} investigated an approach that combined an autoencoder network with an interference detection filter to suppress automotive radar interference. The authors found that using a gated convolution, the encoder network can learn the pattern of the residual signal free of interference. In the decoder network, the interfering portion of the signal can then be recovered. Their results showed that the method improved SINR while also ensuring robustness of the radar measurements in scenarios that have significant interference not represented in the training dataset. Fig. \ref{fig:ae_example}, taken from \cite{chen2021dnn}, shows a plot of the SINR (in $dB$) against the percentage of discarded samples that compares different interference mitigation techniques such as zeroing, auto-regressive (AR) models, autoencoders and the original signals. We see that at 15\% (or greater) of discarded samples, the SINR of both AR and zeroing methods decreased while the autoencoder has a more robust performance.

\begin{figure}[!bt]
\centering
\includegraphics[scale=0.9]{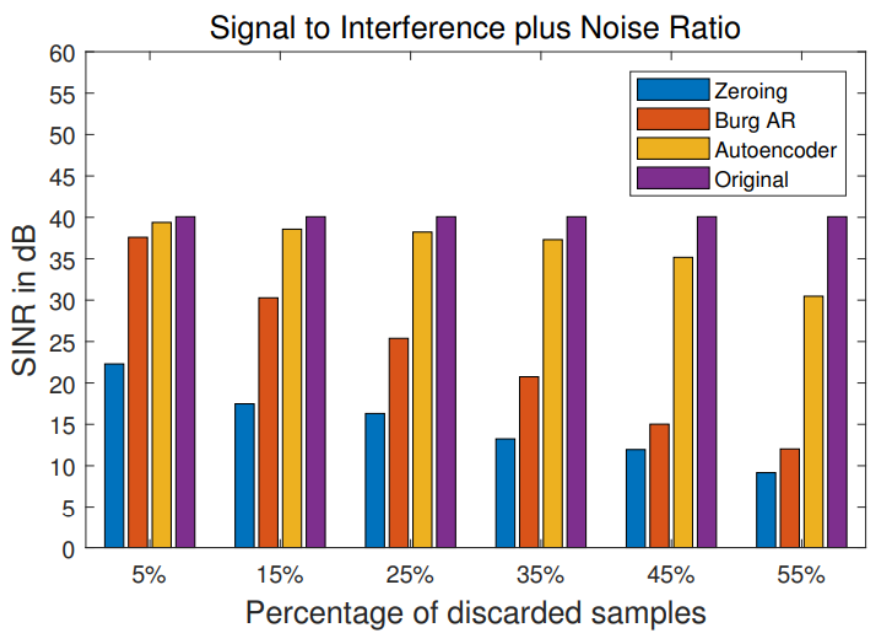}
\caption{Comparison of Autoencoder-based Interference Mitigation technique with Zeroing and Auto-regressive models. The figure was taken from paper \cite{chen2021dnn} titled: \textit{A DNN Autoencoder for Automotive Radar Interference Mitigation}; published in IEEE International Conference on Acoustics, Speech and Signal Processing (ICASSP).}
\label{fig:ae_example}
\vspace{-0.2in}
\end{figure}

Channel autoencoder-based techniques have also been used in joint optimization of transmitters and receivers to suppress interference \cite{dorner2017deep,o2017introduction,erpek2018learning,wu2020deep,wu2019adaptive}. For instance, a channel autoencoder-based interference suppression technique where an encoder and a decoder were jointly optimized in the presence of an interfering signal in a MIMO communication system was investigated by Erpek \textit{et al.} \cite{erpek2018learning}. The authors considered a Gaussian co-channel interference scenario, where the transmitters and the receivers use the same channel. Their technique exploited the structure of the interfering signal and then minimized symbol error rate (SER). Their results show that for a $2 \times 2$ system at SINR higher than $16$ dB, the AE-based model performs significantly better than the baseline (single user single input single output (SISO) systems using QPSK modulation with no interference and no channel coding). One key benefit of jointly training the transceiver in the autoencoder system is that the receivers learn to suppress interference and perform better in comparison to baseline techniques. Despite the fact that the baseline approach may perform better when channel coding is utilized, the autoencoder based suppression scheme from \cite{erpek2018learning} still significantly performed better than the uncoded conventional communication system because it removes the interference. The authors also showed that the autoencoder performance will continue to improve as the network is trained longer at every SINR level.

\subsection{Contributions for Interference Suppression Using DNNs}
Next, we look at deep learning techniques that utilized deep neural networks (DNN) to suppress interference. Typically, many other deep learning algorithms can be regarded as subcategories of DNNs. In this section we consider neural networks, without a convolutional structure, that have at least 4 layers with the inclusion of the input and output layers.  

DNN-based methods are particularly useful in self-interference cancellation (SIC) applications since they can successfully learn a large number of parameters with a much lower computation complexity. For instance, Muranov \textit{et al.} \cite{muranov2021deep} proposed two DNN-based methods - time-invariant \& time-varying non-linear distortion - to mitigate non-linear SI in full duplex relay link. For both methods, they utilized a feed-forward DNN that has two hidden layers and ReLU activation function. For the former, they utilized a DNN to estimate to estimate the non-linear distortion that does not vary with time and then utilized a least squares method to estimate the SI for a time varying channel. For the latter method, they used a transfer learning based DNN to estimate the time varying non-linear distortion. First, they decompose the SI signal into its linear and non-linear components and then estimated and subtracted the linear component from the SI which is then fed to the DNN. Fig. \ref{fig:sic} shows a comparison between the investigated approaches (TID and TVD) and other methods used for SIC. We note that the DNN-based canceller shown in Fig. \ref{fig:sic} proposed by \cite{balatsoukas2018non} is another non-linear interference cancellation technique that was developed to outperform non-deep learning non-linear cancellation techniques that relied on polynomial bases functions. Similar to \cite{muranov2021deep}, DNN was used to construct the non-linear component of the digital signal. The major difference between \cite{muranov2021deep} and \cite{balatsoukas2018non} is that the latter utilized only one hidden layer and thus less complex than the former. This is not surprising since the goal of the \cite{balatsoukas2018non} was to improve computation complexity in comparison to non-DNN techniques like polynomial-based cancellers. In \cite{kurzo2018design}, the authors investigated a hardware implementation of the SIC technique proposed in \cite{balatsoukas2018non}.

\begin{figure}[!bt]
\centering
\includegraphics[scale=0.825]{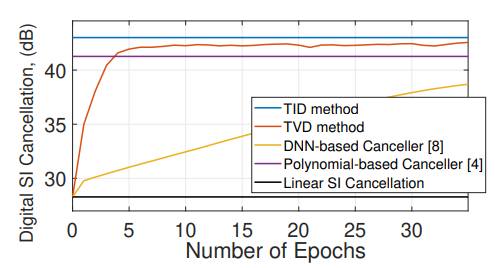}
\caption{Comparison of two DNN-based Self Interference Suppression Techniques with another DNN-based SIC technique and other non-DNN based techniques. The figure was taken from paper \cite{muranov2021deep} titled: \textit{On Deep Learning Assisted Self-Interference
Estimation in a Full-Duplex Relay Link}; published in IEEE Wireless Communications Letters.}
\label{fig:sic}
\vspace{-0.2in}
\end{figure}

In another effort to suppress SIC, Guo \textit{et al.} \cite{guo2019dsic} utilized a DNN technique to cancel non-linear SI. They collected channel data using an SDR testbed consisting of universal software radio peripheral (USRP) in different scenarios transmitting images, texts and numbers. Their DNn structure consist of two hidden layers each followed by ReLU activation function layers. Simply put, their approach showed that their DNN model is stable, fast and can successfully cancel SI with a high accuracy. Shi \textit{et al.} investigated a feed forward NN to reconstruct SI signal. First, an adaptive linear filter was used to extract the transmitted baseband signal features and then the features were utilized to train the NN. Similar to \cite{balatsoukas2018non}, their feed forward NN consist of one hidden layer and suffers from a similar shortcoming. It is important to note that while DNN techniques outperform non-DNN techniques in SI suppression across different channel environment and modulation systems, we also observed that other DNN-based SI suppression techniques either reduce computation complexity or require fewer hardware resources and may not necessarily outperform polynomial-based SI suppression techniques. 

Another paper that utilized a DNN for interference suppression is \cite{xu2018deep}. In this paper, Xu et. al. \cite{xu2018deep} developed a technique to cancel interference for non-orthogonal multiple access (NoMA) techniques using a DNN. In OFDM based systems, the sub-carriers are orthogonal to one another, hence the issue of cross interference is prevented. However, OFDM-based techniques usually require that the sub-carriers are spaced and the cyclic prefix (CP) is used to ensure that the receiver accurately demodulates the signal. This can sometimes impact spectral efficiency \cite{farhang2011ofdm} since some parts of the much needed bandwidth is required to accommodate the overhead of adding the CPs. On the other hand, NoMA algorithms such as the spectrally efficient frequency division multiplexing (SEFDM) waveform utilize non-orthogonal sub-carriers, hence leading to interference from adjacent sub-carriers \cite{ghannam2018sefdm}. In \cite{xu2018deep}, their neural network consisted of four layers including two hidden layers. Compared to some of the CNN algorithms described above that utilized the ReLU activation function, the authors in \cite{xu2018deep} utilized a sigmoid activation function at each layer. They trained a DNN algorithm offline with $40k$ QAM symbols generated using simulated random data. In evaluating the performance of their DNN, they utilized the bit error rate (BER)\footnote{Mathematically, BER can be defined as a ratio of errors to the total number of bits. When a signal is strong, the corresponding BER is small while the converse is true when a signal is weak.} metric for comparison. Although theirs is simulation-based investigation, their key conclusion is that deep neural networks can suppress inter carrier interference (ICI) within SEFDM signal waveform. Also, they achieved a significant gain in SNR when compared to conventional techniques that utilize hard decision detectors. Another key findings is that the neural network performance depends strongly on the waveform characteristics. For instance, the first neural network considered in \cite{xu2018deep} is the \textit{no connection-neural network (N-NN)} in which each symbol in one sub-carrier is an independent network and the number of sub-networks is determined by the quantity of sub-carriers in the SEFDM signals. In another neural network proposed--\textit{partial connection-neural network (P-NN)}--two symbols are connected in one sub-net. Other neural network architectures are discussed in \cite{xu2018deep}. 

In another DNN-based interference suppression approach, Hager \textit{et al.} \cite{hager2018nonlinear} investigated a learned digital backpropagation (LDBP) technique to mitigate non-linear interference for a fiber optic communication application. The designed network is based on unrolling the split-step Fourier method (SSFM)\footnote{SSFM is a numerical method to implement DBP \cite{hager2018nonlinear}.} that has a similar functionality as a DNN \cite{hager2018nonlinear}. In the LDBP approach, the SSFM serves as a basis for a complex-valued DNN and then is further used to optimize the network parameters. The network parameters include the weights, attenuation, dispersion, and non-linearity parameters. Their results show that the LDBP achieves a 50\% complexity reduction when compared to the DBP. Also, one key benefit to using this approach is that the LDBP method helps in choosing the appropriate hyper-parameter choices (e.g., number of network layers, type of non-linearity, etc.) that usually plagues DNN algorithm adoption.  Compared to standard “black-box” deep NNs, this approach leads to clear hyper-parameter choices. 

Further, Promsuk \textit{et al.} \cite{promsuk2021development} proposed a multi-layer perceptron (MLP) algorithm to reduce ACI in an IoT network. In their proposed model, they utilized a FFT data and amplitude as input to the MLP model. The input signals were binary phase shift keying (BPSK) and 16QAM modulated signals that were synthetically generated with path loss and small scale fading to mimic a more realistic IoT network environment. To mitigate the ACI, they combined a MLP with a minimum mean square error (MMSE) filter. The output from the model when the input data is the amplitude is the estimated amplitude of the interference signal. The estimated amplitude of the received signal is then evaluated using a system of equation used in \cite{promsuk2021development}. On the other hand, with an FFT input signal, the output is the estimated FFT of the interference signal. An inverse FFT of the signal is then computed. Their results showed that the performance of the MLP model for both types of input data outperforms the method that utilized the MMSE alone in terms of accuracy, precision, recall and F1-scores.  

In another effort, Mosavi \textit{et al.} \cite{mosavi2016narrowband} investigated a DNN interference suppression technique for mitigating narrow-band interference for global positioning systems (GPS) navigation application. Their method leverages adaptive notch filters as a cascade filter and a simple DNN structure. The DNN structure can be trained quickly because the adaptive notch filters are used instead of updating weights and threshold typically done in MLP models. This technique also reduces exponential computation incurred while updating weights and thus makes it attractive for hardware implementation \cite{mosavi2016narrowband}. Their simulation results showed that they were able to improve SNR of the intended receiver in the presence of single-tone continuous wave interference (SCWI) and multi-tone continuous wave interference (MCWI) when compared to MLP, RNN and other compound methods. 

Finally, Shi \textit{et al.} \cite{shi2012interference} proposed an approach to suppress interference in a direct sequence spread spectrum (DSSS) system using a combination of a power distributing predominance wavelet packet transform (PDP-WPT) and an extended backpropagation neural network (EBPNN) in a satellite navigation system. The process involves a step that first tracks the interfering signal effectively and then identifies the transmitted signal using the EBPNN based on its structural simplicity. Their results show that the proposed method can eradicate strong interference in the received signal with a good signal correlation. Also, the mean square error (MSE) of the proposed technique is smaller compared with other conventional techniques. Another benefit of the technique is its higher convergence rate when compared to other time-domain interference suppression methods \cite{shi2012interference}.

In summary, deep neural networks can be used to model and extract interference. One major benefit of DNNs over analytical models is that they can easily adjust to real-world imperfections \cite{hager2018nonlinear}. Next, we compare the benefits and drawbacks of each deep learning techniques. This comparison helps the reader to gain some insights into why deep learning techniques, in general, are being adopted in interference suppression and what challenges may be associated with using them.

\section{Comparison Between the Approaches}\label{sec:comparison}
Table \ref{table:DL Comparison} contains a summary of the DL-based approaches used by existing works in the interference suppression literature. We look at features such as application/type of interference, deep learning methods used, the type of dataset and the performance metrics used for evaluation. The interference types considered include radar, optical, and a variety of wireless communication standards, amongst others. From our findings, there are more papers on CNN for interference rejection or suppression compared to other deep learning techniques. 

\subsection{Application/Interference Type}
In papers where the type of interference considered is not explicitly stated we provide the application considered (for example \cite{xu2018deep} with optical communication systems). Other interference types considered include heterogeneous technology interference between different wireless technologies sharing the same frequency band such as those of \cite{lin2020cross,zheng2020mitigating}. Some of the most popular applications where interference suppression was studied is the automotive radar applications such as \cite{ristea2020fully,fuchs2020automotive, rock2019complex, rock2021resource, fuchs2021complex}.

\subsection{Deep Learning Algorithm}
The deep learning algorithms range from supervised learning techniques such as vanilla deep neural networks (DNN), CNNs, RNNs; to unsupervised learning techniques such as autoencoders that utilize a network of encoders and decoders to learn salient features of interfering signals from input data and then consequently removing them from the signals. It is not uncommon to train the encoder or decoder using supervised algorithms such as CNN as done by \cite{fuchs2020automotive}. It is important to point out that using deep learning techniques usually require a lot of data to be able to perform well. For some of these works \cite{lin2020cross, fuchs2020automotive, rock2019complex}, however, the quantity of the datasets used is relatively small when compared to the recommendations from sample complexity study in \cite{oyedare2019estimating}. 

\subsection{Dataset Type}
We also investigate the dataset types used for these tasks. Some works utilized OTA data collected from real hardware transmitters and receivers in the wild. Some other works also utilized simulated datasets generated from MATLAB, NS3 and other simulation engines. Other researchers utilized synthetic data which sometimes could be a combination of simulated and real devices. Some papers \cite{rock2021resource,chen2021dnn} also utilized simulation data in training and then tested with actual OTA data. Although there are numerous sound simulation models, the assumptions made by their creators do not always hold for OTA transmissions in the wild. This negatively impacts the acceptability of their results by other experts in the community.

\begin{table*}[!p]
\caption{Summary of Selected DL-Based Interference Suppression Techniques}
\centering
\footnotesize
\begin{tabular}{|c|p{3cm}|p{2.5cm}|p{1.5cm}|p{2cm}|p{3.5cm}|}
\hline
\textbf{References} & \textbf{Application/ \textbf{interference type}}   & \textbf{DL Algorithm} & \textbf{Dataset type}  &  \textbf{Dataset Size} & \textbf{Performance Metric} \\
\hline
 \cite{ristea2020fully} & FMCW radars & CNN & Simulation & $48k$ & SNR, area under the receiver operating characteristics curve (AUC), mean absolute error (MAE) in dB \\
 \hline
\cite{rock2019complex} &  FMCW radars/Chirp Sequence (CS) & CNN & Simulation & 2,250 scenarios & SINR; EVM \\
\hline
\cite{rock2021resource} & FMCW radars/CS & CNN & Simulation \& OTA $+$ simulated interference & $3k$ range Doppler (RD) maps & F1 Score; RD magnitude spectra; SNR \\
\hline
\cite{fuchs2021complex} & FMCW radars/CS & CNN & OTA $+$ simulated interference & $2.5k$ range Doppler (RD) maps & F1 Score; EVM; PPMSE \\
\hline
\cite{xu2018deep} & Optical Communication Systems & DNN & Simulation & $40k$ & Bit error rate (BER) \\
\hline
\cite{hager2018nonlinear} & Non-linear interference in optical communication systems & Learned DBP & Simulation & \textit{Unknown} & Transmit power vs Q-factor \\
\hline
\cite{promsuk2021development} & IoT Network & Artificial Neural Network - Multi-layer Perceptron (MLP) & Simulation & \textit{Unknown} & CDF of BER \\
\hline
\cite{mosavi2016narrowband} & Narrow-band interference for GPS navigation application & DNN & Simulation & $62.854k$ test samples & SNR; RMS \\
\hline
\cite{shi2012interference} & Direct sequence spread spectrum (DSSS) system & DNN - Extended backpropagation NN (EBPNN) & Simulation & \textit{Unknown} & BER; MSE; Correlation value \\
\hline
\cite{muranov2021deep} & Full Duplex Relay Link & DNN & Simulation & $8k$ (training data size) & BER, SNR \\
\hline
\cite{guo2019dsic} & In-band Full Duplex Wireless & DNN & Simulation \& OTA &  \textit{Unknown} & MSE, BER \\
\hline
\cite{shi2019digital} & Full Duplex Wireless Communication Systems & Feed Forward Neural Network & Simulation & \textit{Unknown} & PSD \\
\hline
\cite{kurzo2018design} & Full Duplex Radios & Neural Network & Hardware Experimentation & \textit{Unknown} & Throughput, Efficiency, Si Cancellation in dB \\
\hline
\cite{yang2010iterative} & Direct Sequence CDMA (DS-CDMA) & Feed Forward Neural Network & Simulation & \textit{Unknown} & BER, SINR, SNR\\
\hline
\cite{mun2018deep} & Time domain interference for chirp sequence radars & Gated Recurrent Units (GRU), recurrent neural network (RNN) & Simulation & $150k$ & Signal to remaining interference noise ratio (SRINR) \\
\hline
\cite{mun2020automotive} & Triangle FMCW; CS; multiple frequency shift keying (MFSK); OFDM radar interference & Attention RNN & Simulation & $150k$ & SINR; 2D- FFT \\
\hline
\cite{wei2016gps} & Single-tone CWI; Multi-tone CWI (MCWI); Pulsed CWI, Linear FMCW & RNN & Simulation & \textit{Unknown} & SNR Improvement; mean square prediction error (MSPE) \\
\hline
\cite{lin2020cross} & Heterogeneous technology interference (Wi-Fi, ZigBee, LTE-U) & Denoising Autoencoders (DAE) & OTA, synthetic (GAN augmented data)  & $8.7k$ & Symbol error rate (SER), signal-to-noise ratio (SNR), interference power (dBm) \\
\hline
\cite{fuchs2020automotive} & FMCW radars & Autoencoder-based convolutional neural network & OTA & $8.78k$ & Error vector magnitude (EVM), signal-to-noise-plus interference ratio (SINR), absolute \& phase structural similarity index (SSIM) \\
\hline
\cite{chen2021dnn} & FMCW/CS for automotive radar applications & Autoencoder & Simulated training data; real OTA data for validation data & $2k$ & SINR \\
\hline
\cite{lynggaard2018using} & Wireless sensor networks & Linear regression (LR) without regularization & Simulation & \textit{Unknown} & Classifier score; packet receive ratio (PRR); battery power savings\\
\hline
\end{tabular}
\label{table:DL Comparison}
\end{table*}

\subsection{Performance Metrics}
Performance metrics used include bit error rate (BER), symbol error rate (SER), SNR, SINR, EVM, signal-to-remaining-interference-noise ratio (SRINR), interference power, packet receive ratio (PRR), F1 score, PPMSE, mean square prediction error (MSPE) amongst others. Xu et. al. \cite{xu2018deep} utilized a plot of the BER with optical SNR (OSNR) to evaluate their interference cancellation technique. A comparison was made between the conventional hard decision technique and a variation of deep neural networks algorithms. For BER, a lot of data would need to be processed before having a single error since practical BERs must be very low ($10^{-1}$ to $10^{-9}$) and $10$-$100$ errors are required for confidence. It is often used in computer simulations to evaluate the performance of digital communication systems \cite{jeruchim1984techniques}. Similar to BER, SER is usually plotted against some measure of signal-to-noise ratio (SNR). SER is the ratio of the number of symbols in error to the total number of symbols transmitted. While SNR is a metric that is used to parameterize the performance of a signal processing system that uses a Gaussian distributed noise \cite{johnson2006signal}. This is a very intuitive way to measure the performance of a digital communication system. For instance, Lin et. al. \cite{lin2020cross} plotted the SER with respect to interference power in order to compare the performance of their denoising autoencoder. In understanding how deep learning can effectively suppress an interfering signal, \cite{lin2020cross} compared corrupted Wi-Fi signal before suppression and the signal after suppression by looking at a plot of the SNR versus the interference power. The resultant signal post suppression can be compared to a clean signal (interference-free) and the neural network output SNR can be very useful in understanding the impact of such a suppression approach.  SINR defines the ratio of the power of a signal of interest to the sum of the interference power and the power of the background noise \cite{haenggi2009stochastic}. For radar systems application, the ratio of the power of a single object reflection to the average remaining interference power after cancellation is referred to as signal to remaining interference plus noise ratio (SRINR) \cite{wagner2018threshold}. Mun et. al. \cite{mun2018deep} utilized this metric to compare the performance of the RNN-based automotive interference suppression technique. 

We provide some details on other common performance metrics used in deep learning literature which are not typically used as wireless communication metrics, yet may provide utility for evaluating interference suppression.  



%


\subsubsection{Area under the receiver operating characteristics curve (AUC)}
The AUC metric is typically used to ascertain detection or classification performance for a multi-class classification problem at various threshold settings. The receiver operating curve (ROC) is a probability curve while the area under the curve (AUC) defines the degree or measure of separability \cite{narkhede2018understanding}. In essence, it describes how well a model can distinguish between classes. It is a plot of the true positive rate (TPR) against the false positive rate (FPR). The higher the AUC, the better the model is at distinguishing between classes. Ristea et. al. \cite{ristea2020fully} used this metric to compare the performance of their suppression technique with other conventional suppression techniques. In their specific application, the closer the value was to $1$ the better. When compared to non-deep learning technique such as zeroing, their deep fully convolutional neural network performed better for both the validation and test sets. 

\subsubsection{Mean Absolute Error (MAE)}
MAE is another simple performance metric that measures the error between the values of an actual quantity and the predicted quantity. For instance, \cite{ristea2020fully} computed the MAE in dB between the range profile amplitude of targets computed from label signals and the amplitude of target from predicted signals. Similar to SNR and AUC, \cite{ristea2020fully} used MAE to compare between their technique and other suppression techniques. Hence, this metric can be applied at the application level.

\subsubsection{Error Vector Magnitude (EVM)}
The EVM is the root mean square (RMS) average amplitude of the error vector (\textit{a vector in the I/Q plane between the ideal constellation point and the point received by the receiver} \cite{wiki:Error_vector_magnitude}) that is normalized to the ideal signal amplitude reference \cite{schmogrow2011error}. Fuchs et. al. \cite{fuchs2020automotive} used the EVM to compare the detection in the original image with the predicted image. Rock et. al. \cite{rock2019complex} also used the EVM to compare the performance of their scheme with the state-of-the-art. This metric is thus a metric that is applicable to the I/Q data.

\subsubsection{Structural Similarity (SSIM) Index}
The SSIM index is typically used for qualitative visual assessment. Research has shown that they perform better than mean squared error (MSE) \cite{wang2004image}. According to \cite{fuchs2020automotive}, it compares image characteristics such as luminance, contrast and structure which are useful when removing interference since images may preserve target peaks resulting from interference. Hence, Fuchs et. al. \cite{fuchs2020automotive} utilized this technique to compare the spectrum free of interference with the reconstructed spectrum after interference removal. Hence, if the output can be viewed as a two dimensional array of I/Q or application output, it is applicable.

\subsection{Benefits and Drawbacks of Deep Learning-based Interference suppression Schemes}
In this section, we compare the benefits and the drawbacks of the deep learning-based interference suppression approaches. This is crucial because despite the gains brought about by applying deep learning to interference suppression, there are some drawbacks. This is also important as it helps the reader to understand what the DL algorithm solves as well as gaining a better understanding of which algorithm may fit their need. Details of the benefits and drawbacks of each of the DL-based suppression techniques can be found in Table \ref{table:Adv_Diasdv}

\section{Open Challenges}\label{sec:challenges}
Despite the successes with using deep learning algorithms in suppressing interference, there are still some challenges that need to be overcome before they can be adopted in the production environment. We note that some of these challenges also hold true for other deep learning-based wireless communication tasks. 

\subsection{Lack of Interpretability of Deep Learning Algorithms}
One of the major challenges with using deep learning to classify interference is their lack of interpretability. Interpretability is a concept that deals with the model transparency and its ability to explain the intricacies of the model after the fact \cite{pulkkinen2020understanding}. As expected, this is very important in suppressing interference sources since this can be useful in subsequently identifying spectrum offenders or devices misusing the spectrum in an opportunistic spectrum access paradigm such as can be found in the  industrial, scientific, and medical (ISM) radio bands. This implies the need for interpretable artificial intelligence (AI). If the deep learning engine is viewed as a black box, it will be difficult for end users to trust the system. As researchers make progress in applying deep learning to interference suppression, privacy concerns may start to grow as we have seen in other deep learning applications \cite{mireshghallah2020privacy,liu2020privacy}.

\subsection{The Stochasticity of the Wireless Channel}
\label{channel}
Another issue with using deep learning in interference suppression is the issue of the randomness of the wireless channels. Most of the papers reviewed in this survey have used data trained in only one type of environment (e.g. an-echoic, AWGN, etc). Meanwhile, in practical deployments, test signals go through a variation in channel environments which can result in the incapability of the trained model to adapt to new test cases. Since, there are numerous channel environments at any given time, it is important to be able to interpret deep learning models so that they can better learn different RF environments. In other words, any difference between training and testing environment could impact the performance of deep learning algorithm. Developing effective GAN approaches for training data creation could be very valuable. Given that deep learning algorithms require huge amount of data to train them, the nexus between the quality and quantity of data results in another challenge that impairs the use of deep learning in interference suppression. Training data collection is very laborious and can be more difficult when a supervised learning algorithm that requires labeling the data is considered.

Often times, researchers use data augmentation techniques to further increase the quantity of their data. For instance, in the image processing domain, one way to improve the quantity of training data is to augment the data using both simple and complex techniques. Simple transformation techniques include flipping, cropping or rotating the images to increase the quantity of such data. Since these techniques are not transferable to I/Q data used in interference suppression, complex techniques that leverage popular generative models such as generative adversarial networks (GAN) and variational autoencoders (VAE) have been proposed to augment the input data image in such a way as to generate new data that belong to the same distribution as the training data but not exactly in the training data. In most data augmentation for wireless communications tasks (including interference suppression, emitter and device classifications), such augmentation techniques have had very little success since the algorithm often times learn noise and other unwanted artefacts in the signal. In addition, the stochasticity of the RF environment makes it difficult to augment datasets for wireless communication applications.

\begin{landscape}
\begin{table}[!hpt]
\caption{Benefits and Drawbacks of DL-Based Interference Suppression Approaches}
\centering
\footnotesize
\begin{tabular}{|p{2.5cm}|p{1.75cm}|p{6cm}|p{6cm}|p{5.5cm}|}
\hline
\textbf{DL Algorithm} & \textbf{References} &\textbf{Technical Approach} & \textbf{Benefits} & \textbf{Drawbacks} \\
\hline
Convolutional neural network (CNN) & \cite{ristea2020fully,rock2019complex,rock2021resource,fuchs2021complex} & \begin{itemize}
    \item Uncorrelated interference is learnt via FCN \cite{ristea2020fully}. \item STFT can be used to convert time domain signal into a spectrogram \cite{ristea2020fully}. \item CNN is used to denoising and thus suppress interference using both interfered data and clean data. \item SIR \& SNR can be used to scale interference and noise powers in relation to the object signal power\cite{rock2019complex}.
\end{itemize} & \begin{itemize}
    \item The CNN algorithms both outperformed the conventional suppression method such as zeroing, ramp filtering and IMAT techniques \cite{rock2019complex}. \item Very useful for learning when the amount of data is relatively smaller. \item Similar to Autoencoders, CNN can be used for denoising \cite{rock2019complex}.
\end{itemize} &  \begin{itemize}
    \item Could result in very high EVM performance when compared to zeroing, ramp filtering and and IMAT). \item High EVM could also result in distortions for radar applications \cite{rock2019complex}. \item Usually require high memory to train large datasets. 
\end{itemize} \\
\hline
Deep neural network (DNN) & \cite{xu2018deep,hager2018nonlinear,promsuk2021development,mosavi2016narrowband,shi2012interference,muranov2021deep,guo2019dsic,shi2019digital,kurzo2018design,zhang2018self} & \begin{itemize}\item DNNs can be useful in cancelling interference for NoMA systems \cite{xu2018deep,ghannam2018sefdm}. \item MLPs can be combined with filters to suppress interference \cite{promsuk2021development}. \item NoMA algorithms such as the SEFDM waveform utilize non-orthogonal sub-carriers which can result in ACI. \cite{ghannam2018sefdm}. 
\item Performance depends on waveform characteristics \item They are usually structurally simpler when compared to other deep learning techniques. 
\end{itemize}
& \begin{itemize}
    \item Useful in mitigating non-linear interference and reducing complexity \cite{hager2018nonlinear} \item DNNs can suppress inter carrier interference (ICI) within certain signal waveform e.g. SEFDM used in \cite{xu2018deep}.\item Capable of learning a huge number of parameters in suppressing SI. \item They can be used to reduce computation complexity for non-linear interference suppression. \item Gains in SNR when compared to techniques that utilize hard decision detectors.
\end{itemize} & \begin{itemize}
   \item High computational complexity \item Simulation-based studies that utilize DNNs are very susceptible to impractical assumptions. \item Could result in smaller MSE compared to conventional techniques \cite{shi2012interference}. \item Varying complexity (such as number of hidden layers, number of parameters, etc) in the DNN model could result in significant performance degradation.
\end{itemize}\\
\hline
Recurrent neural network (RNN) & \cite{mun2018deep, mun2020automotive,wei2016gps,chang1999narrow,xu2008narrowband}& \begin{itemize}
     \item The dynamics of sequences can be captured with RNNs using a network of nodes \cite{lipton2015critical}. \item Interference can be suppressed in time domain for radar systems. \item Transmit signal is reconstructed in the presence of various interference signals. \item They are typically used when input data is temporally correlated (e.g. time domain signals). 
 \end{itemize}  & \begin{itemize}
     \item Low processing time \item Useful for applications where frequency changes linearly.  \item RNNs with attention blocks can capture sequential relationship between time steps. \item Useful in reconstructing original signal. \item RNNs can take in arbitrary input or output sequences. 
 \end{itemize} & \begin{itemize}
     \item RNNs are vulnerable to exploding gradient problem. \item Most of the implementations have only considered simulation data for interference mitigation. \item For some applications, RNN-based methods have slow convergence rate and thus unable to track fast time varying interference \cite{yang2010iterative}.
 \end{itemize} \\
 \hline
Autoencoders (Denoising AE and CNN-based AE) & \cite{lin2020cross,fuchs2020automotive,dorner2017deep,o2017introduction,erpek2018learning,wu2020deep,wu2019adaptive,chen2021dnn} & \begin{itemize}
    \item CNN-based AE is used to de-noise interfered range-Doppler (RD) images \cite{fuchs2020automotive}. \item Autoencoder is used to learn the distribution of the interference component so that it can be suppressed. \item Signal is reconstructed without any knowledge about symbol structure and channel response. \item AE is combined with an interference detection filter to suppress automotive radar interference in \cite{chen2021dnn} 
\end{itemize} & \begin{itemize} 
\item Using real-world data measurement further improves the performance of AE methods.
    \item DAEs perform well when the interfering power is larger than the amplitude of the desired signal. \item Typically, AEs do not not incur any signalling overhead since there's no synchronization between the transmitter, receiver and interferer. \item Signal can be reconstructed without any knowledge about symbol structure and channel response.
\end{itemize}  & \begin{itemize}
    \item AE techniques often perform poorly at low SNR \item AEs perform poorly with insufficient data \item They also incur a huge processing time, hyper-parameter tuning and there is also a need to validate the model.
\end{itemize} \\
\hline
\end{tabular}
\label{table:Adv_Diasdv}
\end{table}
\end{landscape}

\subsection{Performance Analysis}
As we have shown in this survey, many of the researchers utilized different metrics to show the performance of their algorithms. Also, the majority of the DL-based interference suppression techniques utilized I/Q data as input into the algorithm. Compared to rigorous and well established  mathematical foundation of signal processing methods developed by earlier pioneers of the wireless communications field (such as Claude Shannon's information theory mathematical derivations, amongst others), the lack of a robust mathematical underpinning for analyzing results from deep learning algorithms may hinder their practical adoption for interference suppression. On the other hand, it has been shown that training and testing data do not usually belong to the same distribution \cite{liu2021detecting}, hence, information-theoretic techniques may still be explored to help with performance metrics that would be suitable for interference suppression research. Furthermore, metrics such as BER, while ideal, might be slow in evaluating performance since extreme amounts of data would need to be processed for one error.

\subsection{Issues with Open Set Recognition (OSR) or Out-of-Distribution (OOD) Data}
Open set recognition (OSR) is a common phenomenon in which unknown classes not present in the training phase during learning are found in the testing phase \cite{draganov2020open,mattei2019feature}. Also known as out-of-distribution (OOD) data \cite{liu2021detecting}, this issue can result in a lack of trust for deep learning models since the test set in real world applications is infinite, and even more so in wireless communications. For example, we discussed earlier how the random nature of the wireless channel can result in a drastic degradation in the performance of a deep learning classifier and also the fact that the number of possible channels is infinite. Typically, in deep learning-based algorithms, researchers make the assumption that the training and testing data are sampled from the same distribution. As a result of this assumption, the ability of the model to generalize well to unseen testing data is often accounted for. However, when the deep learning algorithm is deployed in real-world applications, testing data is not restricted to the same distribution as the training data. Hence, it is feasible and probable for the testing data to differ significantly from the training data.

Some prior works \cite{nguyen2015deep,moosavi2017universal} have shown that DL models usually make high confidence predictions for inputs that are either irrelevant or cannot be recognized. Although, this phenomenon was briefly mentioned in \cite{mattei2019feature}, we believe that it has not received enough attention in the DL-based interference suppression research. For instance, in safety- and mission-critical applications, such as autonomous driving, OSR or OOD data can lead to perception errors and cause accidents or system failures. Further, the ability to detect and consequently suppress interference using these deep learning techniques could be negatively impacted by OOD or OSR. Another reason why this issue may be consequential is that because they do not result in explicit model errors, they may be undetected. Thus, we believe that in order for deep learning to be safely deployed in interference suppression research, there is an important need to distinguish between the known and unknown classes in the testing data. 

\subsection{Challenges with Implementation}
Training a deep learning algorithm is usually time-consuming and computation-intensive, thus most researchers train the algorithms offline and then deploy a trained algorithm in their application. For instance, Grimaldi \textit{et al.} \cite{grimaldi2018real} observed that an online classifier, distinguishing between Bluetooth (also known as IEEE 802.15.1) and Bluetooth low energy (BLE)\footnote{Bluetooth is a terrestrial, adhoc, short range communication protocol \cite{tjensvold2007comparison}. While BLE was designed for a reduced power consumption and cost with a similar communication range as Bluetooth\cite{gomez2012overview}, it is important to note that Bluetooth and BLE are independent of each other.} technologies is ten times more likely to mis-classify when compared to offline methods which only mis-classified $2\%$ of the time. This occurred because the two technologies have very similar PHY layer with both of them utilizing the Gaussian frequency-shift keying (GFSK)\footnote{GFSK filters the data pulses with a Gaussian filter instead of changing the frequency at the beginning of each symbol period so as to make the transitions smooth. GFSK is used for low power design requirements such as BLE \cite{chang2006digital}.} modulation. Furthermore, Schmidt \textit{et al.} \cite{schmidt2017wireless} ran their multi-layered CNN algorithm using expensive graphical processing units (GPU) servers which were used to train the algorithm, this expensive hardware could pose a barrier to the adoption of deep learning to practical interference suppression. While offline training works well in the computer vision research where noise and other factors do not disproportionately impact performance, the same could not be said for interference suppression application. Since the interference characteristics of the received signal used to train an offline algorithm changes with time, researchers may find that their trained algorithm may not be suitable in different scenarios or channel environment or even at another time period. This would imply that retraining needs to be done during deployment. This retraining may happen more frequently than is done in other applications. For computation-constrained applications, this would quickly result in a bottleneck. One possible solution to this is to implement the deep learning algorithm directly on the deployed hardware so that training time can be more manageable in real-time or near-real-time systems. Again, for resource-constrained IoT devices, this may be infeasible.

\section{Conclusion}\label{sec:conclusion}
In this survey paper, we have identified key characteristics of interference in several wireless communication applications. We have reviewed deep learning techniques that have been utilized by researchers to suppress interference. We also provided a comparison of such techniques for interference suppression. Our findings show that deep learning techniques have been shown to significantly improve interference suppression when compared to traditional suppression techniques. Also, deep learning techniques can learn complex non-linear relationship in the input data, thereby making them useful in learning complex non-linear decision boundaries. We also noted that deep learning techniques require access to a large dataset so that the algorithm can be sufficiently trained thereby improving test performance. It is important to emphasize that the application of deep learning to interference suppression will continue to grow. In conducting this survey, the authors found that deep learning techniques such as autoencoders and CNN are of particular interest, given their unique characteristics that can be leveraged in interference suppression. While there are a lot of merits to using deep learning to suppress interference, we have also found some challenges that need to be addressed to ensure success. One key challenge with using deep learning is the availability of benchmark dataset, the CRAWDAD database is a potent tool that can serve as a reservoir of key datasets that can help in interference suppression. Lastly, we note that there is still a need to further investigate concepts such as explainable artificial intelligence (AI) so we can better understand the inner workings of deep learning models.


\bibliographystyle{ieeetr}
\bibliography{ref}

\begin{IEEEbiography}[{\includegraphics[width=1in,height=1.25in,clip,keepaspectratio]{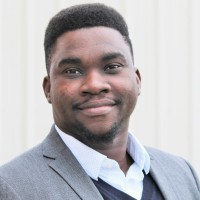}}]{Taiwo Oyedare} received his Bachelor’s degree in Electrical and Electronics Engineering from Ekiti State University, Ado-Ekiti, Nigeria in 2012. He received his Master’s degree in Computer and Information Systems Engineering from Tennessee State University in 2016. Taiwo is currently a PhD student in the department of Electrical and Computer Engineering at Virginia Tech. His research interests include wireless security and the application of deep learning to wireless communications. 
\end{IEEEbiography}

\begin{IEEEbiography}[{\includegraphics[width=1in,height=1.25in,clip,keepaspectratio]{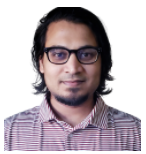}}]{Vijay K. Shah} is an Assistant Professor in the Cybersecurity Engineering (CYSE) Department at George Mason University, VA, USA. Before joining Mason, he was a Research Assistant Professor with the Bradley Department of Electrical and Computer Engineering (ECE) at Virginia Tech, and affiliated with Wireless@Virginia Tech. His research interests include Next-G wireless, O-RAN architecture, Robust AI/ML for communications, and wireless testbed development and prototyping -- with applications to internet of things (IoT) systems, autonomous systems (e.g., V2X) and cyber-physical systems (e.g., power grids). 


\end{IEEEbiography}

\begin{IEEEbiography}[{\includegraphics[width=1in,height=1.25in,clip,keepaspectratio]{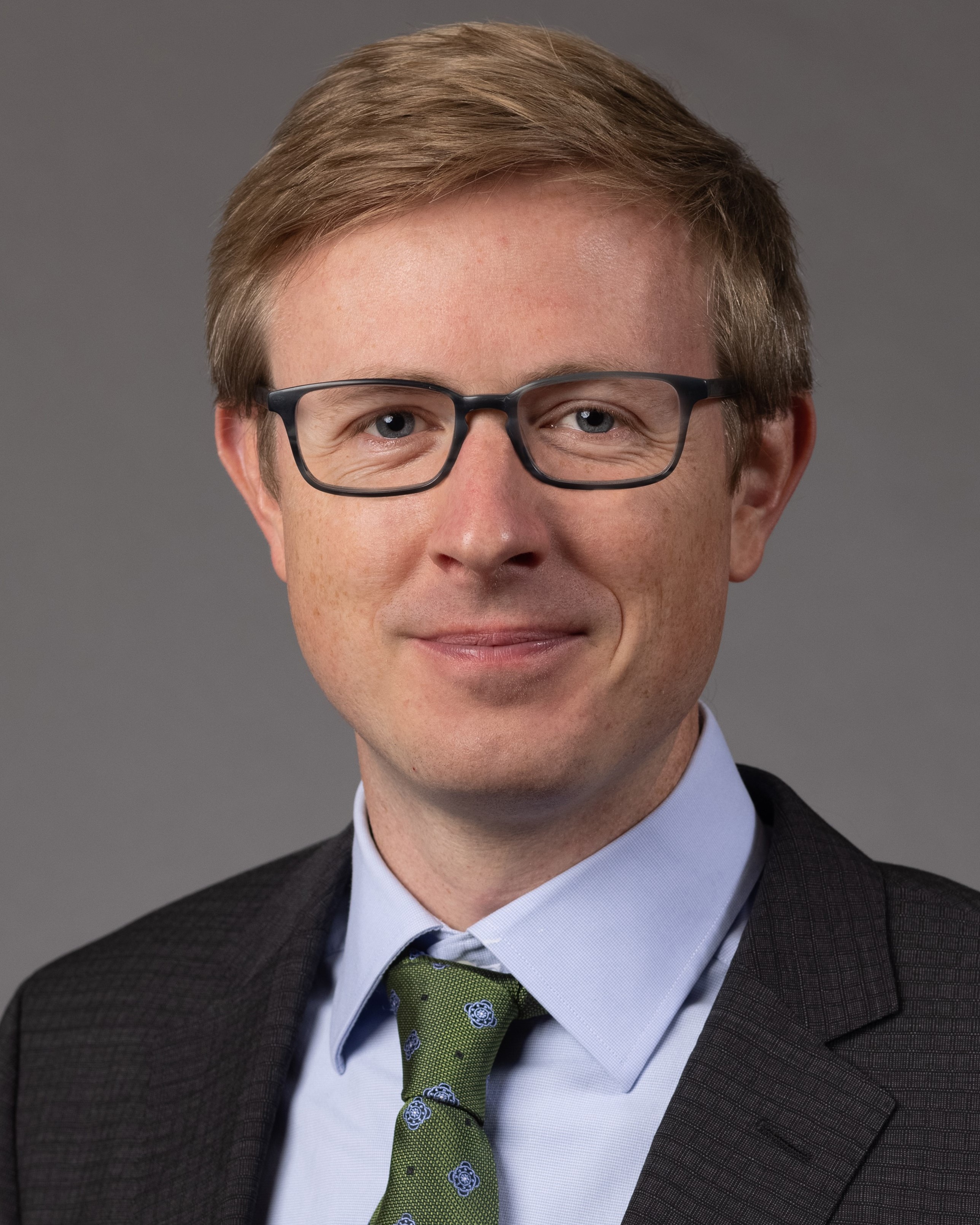}}]{Daniel Jakubisin} is a Research Assistant Professor with the Virginia Tech National Security Institute and with the Bradley Department of Electrical and Computer Engineering at Virginia Tech, by courtesy. His research is directed towards advancing the security and resiliency of wireless communication systems. Prior to joining Virginia Tech, he was a technical advisor with McGuireWoods LLP supporting IP/patent matters in the telecommunications sector. He received his Ph.D. and M.S. degrees from Virginia Tech, Blacksburg, VA, USA and his B.S. degree from UNC Charlotte, Charlotte, NC, USA.
\end{IEEEbiography}

\begin{IEEEbiography}[{\includegraphics[width=1in,height=1.25in,clip,keepaspectratio]{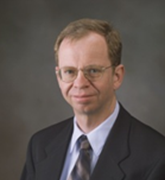}}]{Dr. Jeffrey H. Reed} is the founder of Wireless @ Virginia Tech, and served as its Director until 2014. He is the Founding Faculty member of the Ted and Karyn Hume Center for National Security and Technology and served as its interim Director when founded in 2010. His book, Software Radio: A Modern Approach to Radio Design was published by Prentice Hall and his latest textbook Cellular Communications: A Comprehensive and Practical Guide was published by Wiley-IEEE in 2014. He is co-founder of Cognitive Radio Technologies (CRT), a company commercializing of the cognitive radio technologies; Allied Communications, a company developing spectrum sharing technologies; and for PFP Cybersecurity, a company specializing in security for embedded systems. 

In 2005, Dr. Reed became Fellow to the IEEE for contributions to software radio and communications signal processing and for leadership in engineering education. In 2013 he was awarded the
International Achievement Award by the Wireless Innovations Forum. In 2012 he served on the President’s Council of Advisors of Science and Technology Working Group that examine ways to transition federal spectrum for commercial use. Dr. Reed is a past member CSMAC a group that provides advice to the NTIA on spectrum issues.
\end{IEEEbiography}
\end{document}